\documentclass[twocolumn,english,prd]{revtex4}
\usepackage{graphicx}
\usepackage{bm}
\usepackage{bbm}
\usepackage{amssymb}
\usepackage{amsmath}

\newcommand{\bdm}{\begin{displaymath}}
\newcommand{\edm}{\end{displaymath}}
\newcommand{\Xv}{{\bf X}}

\newcommand{\beq}{\begin{equation}}
\newcommand{\eeq}{\end{equation}}
\newcommand{\enq}{\end{equation}}
\newcommand{\bea}{\begin{eqnarray}}
\newcommand{\eea}{\end{eqnarray}}
\newcommand{\beqa}{\begin{eqnarray}}
\newcommand{\eeqa}{\end{eqnarray}}
\newcommand{\bt}{\begin{tabular}}
\newcommand{\et}{\end{tabular}}
\newcommand{\bdis}{\begin{displaymath}}
\newcommand{\edis}{\end{displaymath}}

\newcommand{\mv}{{\bf m}}
\newcommand{\lv}{{\bf l}}
\newcommand{\kv}{{\bf k}}

\newcommand{\wv}{{\bf w}}
\newcommand{\vv}{{\bf v}}
\newcommand{\uv}{{\bf u}}
\newcommand{\rv}{{\bf r}}
\newcommand{\xv}{{\bf x}}
\newcommand{\yv}{{\bf y}}
\newcommand{\pv}{{\bf p}}
\newcommand{\qv}{{\bf q}}

\newcommand{\Pv}{{\bf P}}

\def\Mpc{\, h^{-1} \, {\rm Mpc}}

\def\kMpc{\, h \, {\rm Mpc}^{-1}}

\def\fun#1#2{\lower3.6pt\vbox{\baselineskip0pt\lineskip.9pt
        \ialign{$\mathsurround=0pt#1\hfill##\hfil$\crcr#2\crcr\sim\crcr}}}
\def\la{\mathrel{\mathpalette\fun <}}
\def\ga{\mathrel{\mathpalette\fun >}}

\begin{document}

%
%
\title{Generation of Vorticity and Velocity Dispersion by Orbit Crossing}
%
%
\author{Sebasti\'an Pueblas \& Rom\'an Scoccimarro}

\vskip 1pc

\affiliation{Center for Cosmology and Particle Physics, \\ 
Department of Physics, New York University, \\
4 Washington Place, New York, NY 10003, USA}

\begin{abstract}
%

We study the generation of vorticity and velocity dispersion by orbit crossing using cosmological numerical simulations, and calculate the backreaction of these effects on the evolution of large-scale density and velocity divergence power spectra. We use Delaunay tessellations to define the velocity field, showing that the power spectra of velocity  divergence and vorticity measured in this way are unbiased and have better noise properties than for standard interpolation methods that deal with mass weighted velocities. We show that high resolution simulations are required  to recover the correct large-scale vorticity power spectrum, while poor resolution can spuriously amplify its amplitude by more than one order of magnitude. We measure the scalar and vector modes of the stress tensor induced by orbit crossing using an adaptive technique, showing that its vector modes lead, when input into the vorticity evolution equation, to the same vorticity power spectrum obtained from the Delaunay method. We incorporate orbit crossing corrections to the evolution of large scale density and velocity fields in perturbation theory by using the measured stress tensor modes. We find that at large scales ($k\simeq 0.1 \kMpc$) vector modes have very little effect  in the density power spectrum, while scalar modes (velocity dispersion) can induce percent level corrections at $z=0$, particularly in the velocity divergence power spectrum.  In addition, we show that the velocity power spectrum is smaller than predicted by linear theory until well into the nonlinear regime, with little contribution from virial velocities. 
\end{abstract}


\maketitle

\section{Introduction}

The evolution of cosmological perturbations is determined, at scales larger than those where baryonic physics becomes important, by the gravitational clustering of cold dark matter, which can be taken as collisionless to a very good approximation.   Therefore, in this regime the Vlasov equation, i.e. the collisionless limit of the Boltzmann equation, describes the dynamics of cosmological perturbations~\cite{1980lssu.book.....P}. 

At large scales, where orbit crossing may be neglected, the Vlasov equation reduces to the dynamics of a pressureless perfect fluid (hereafter PPF). The PPF approximation has been used extensively in analytic approaches such as standard perturbation theory (hereafter PT; see~\cite{2002PhR...367....1B} for a review) and the more recent renormalized perturbation theory~\cite{2006PhRvD..73f3519C} (hereafter RPT) and related techniques~\cite{2007PhRvD..75d3514M,2007A&A...465..725V,2007JCAP...06..026M,2008ApJ...674..617T,2007arXiv0711.2521M,2008arXiv0806.0971P}.

At small scales, as the first nonlinear structures are formed, orbit crossing generates a nontrivial stress tensor (the second cumulant of the phase space distribution function), which leads to velocity dispersion and vorticity in the dark matter distribution. Both of these effects would not be significantly present otherwise; vorticity corresponds to vector modes which are not produced primordially (at least in the simplest models of inflation) and even if they were they decay due to the expansion of the universe, velocity dispersion does get generated primordially e.g. during thermal equilibrium of dark matter in the early universe but for cold dark matter candidates typical values are vanishingly small ($\sim 10^{-6}$ km/s for WIMPs and at most $\sim10^{-10}$ km/s for axions~\cite{1997PhRvD..56.1863S}) compared to typical velocity flows generated during structure formation. 

Although a number of works have attempted to go beyond the PPF approximation~\cite{1993ApJ...416L..71W,1998A&A...335..395B,2000PhRvD..62j3501D,2001A&A...379....8V,2002MNRAS.334..435D,2003ApJ...583L...1S,2005A&A...438..443B}, there has been no quantitative estimate in the literature at what scale corrections to the PPF become important. This is the main goal of this paper.  In order to achieve this, one has to compare PPF solutions with solutions to the Vlasov equation. N-body simulations of collisionless cold dark matter attempt to solve the latter by discretizing the distribution function using particles that follow the characteristics of the Vlasov equation~\footnote{For direct solutions of the Vlasov equation see~\cite{2005MNRAS.359..123A} are references therein.}. The N-body solution will differ from the PPF in regions where particle orbits cross, also known as ``caustics" or ``shell-crossing" in the context of the spherical collapse. This generates a nontrivial stress tensor and higher-order cumulants of the distribution function in the dark matter. We use the N-body solution to measure the induced stress tensor generated by orbit crossing and calculate from it the corrections to the PPF predictions for the density and velocity divergence power spectra at large scales. Recent work on orbit crossing has concentrated on enhancement of dark matter annihilation in caustics~\cite{2006MNRAS.366.1217M,2008MNRAS.385..236V,2008arXiv0809.0497W}. We are instead interested on the impact of orbit crossing on the large-scale dynamics, outside dark matter halos.

Along the way we provide a number of results regarding the nonlinear evolution of peculiar velocities, which compared to the density field has not been studied in as much depth. The generation of vorticity and velocity dispersion impacts the reconstruction of primordial fluctuations from peculiar velocities~\cite{1989ApJ...336L...5B,1993ApJ...412....1D} which assume a cold (single-stream) potential flow, as do other reconstruction methods based on galaxy positions~\cite{1997MNRAS.285..793C,2006MNRAS.365..939M,2007ApJ...664..675E}. The understanding of the nonlinear evolution of the {\em volume weighted} (as opposed to mass or galaxy weighted) velocity field is important, since the velocity difference PDF is one of the building blocks that contributes to the redshift space galaxy power spectrum, independent of galaxy bias and calculable from first principles (\cite{2004PhRvD..70h3007S}, see also~\cite{2008arXiv0808.0003P} for recent discussion). The study of the peculiar velocity field has been recently highlighted as a means to constrain dark energy and large-distance modifications of gravity~\cite{2007PhRvD..76j3523F,2008arXiv0802.3897S,2008arXiv0807.0810S}. Finally, as old and new techniques to measure the peculiar velocity power spectrum improve, some of the issues we study here should be important for making predictions that model nonlinear effects accurately for future observations~\cite{1995PhR...261..271S,2007MNRAS.379..343W,2008JCAP...02..022H,2007PhRvL..99h1301G,2008MNRAS.388..884Z,2008arXiv0802.1935A}.

This paper is organized as follows. 
In section~\ref{secdelresults}, we present results for the power spectrum of velocity divergence and vorticity that follow from applying the Delaunay method to our N-body simulations.   
We discuss how vorticity and velocity dispersion get generated by orbit crossing in section~\ref{orbitcross}, where we also propose an estimator of the stress tensor induced by orbit crossing based on an adaptive method.  In section~\ref{PPFcorrect} we extend PT to include vorticity and velocity dispersion. Finally we present our conclusions in section~\ref{conclude}.

\begin{figure}
\begin{center}
\includegraphics[width=75mm]{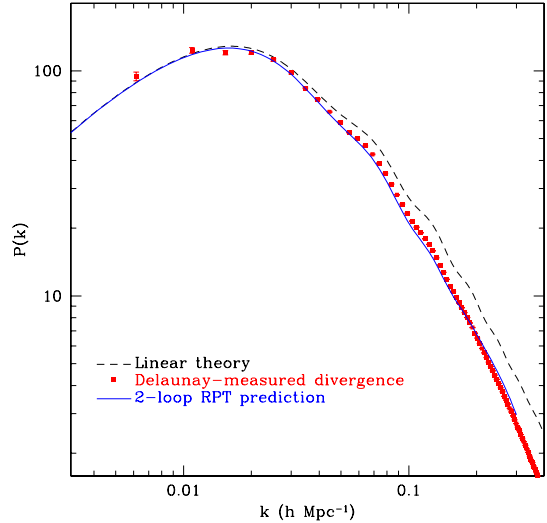}
\end{center}
\caption{Velocity divergence power spectrum at $z=0$ from 50 realizations of the LR1280 simulations. The symbols with error bars denote the velocity divergence power spectrum measured with the Delaunay method, normalized as in Eq.~(\ref{norm}). The  blue solid line is the RPT prediction, and the dotted line is the linear power  spectrum.}
\label{rptvsdel}
\end{figure}

\begin{figure*}
\begin{center}
\hspace*{-20mm}
\includegraphics[width=200mm]{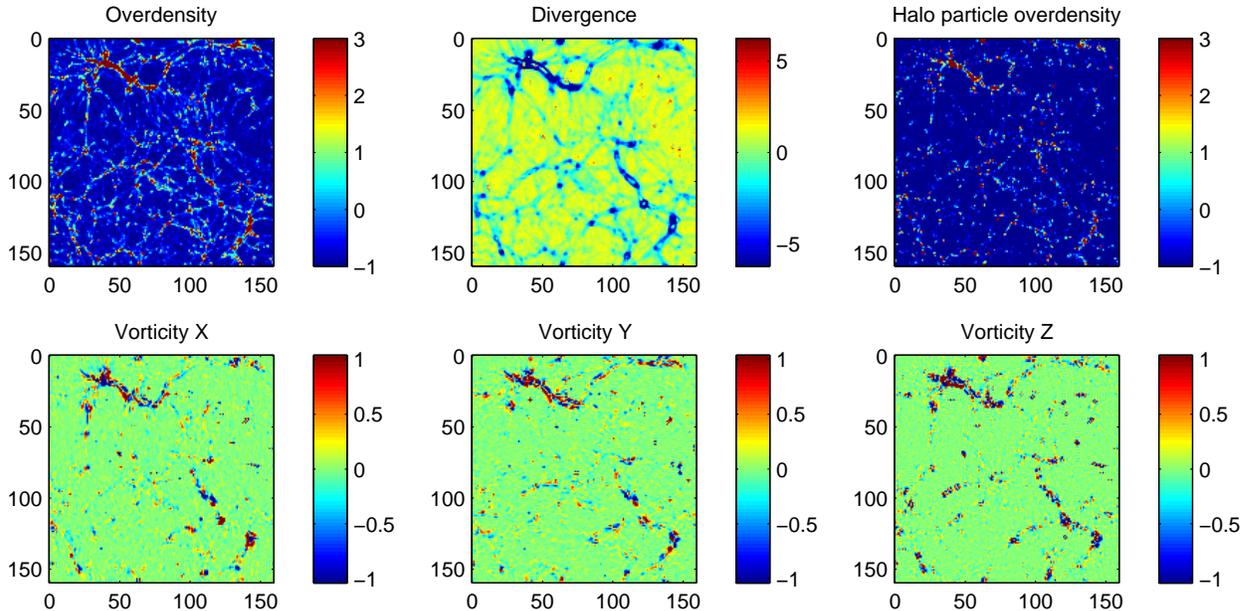}
\end{center}
\vspace*{-10mm}
\caption{Illustration of overdensity, divergence and vorticity in a  $1 \Mpc$ thick cross-section of the simulation box at $z=0$. The divergence and   vorticity components panels correspond to the dimensionless quantities  $\nabla\cdot\uv/{\cal H}f$ and $\nabla\times\uv/{\cal H}f$. The panel labeled ``Halo particle overdensity'' shows the overdensity  of particles belonging to dark matter halos with mass $m \protect\ga 2.4 \times 10^{10} h^{-1}M_\odot$.}
\label{vortilus}
\end{figure*}

\section{Divergence and Vorticity Power Spectra}
\label{secdelresults}

\subsection{Spatial Distribution in N-Body Simulations}

In Appendix~\ref{secdelaunay} we discuss how to estimate the velocity field from  Delaunay tessellations, also comparing to more standard interpolation methods that deal with mass weighted velocities. We refer the reader to the Appendix for technical details. Here we show the results of applying the Delaunay method to estimate velocity statistics from the cosmological simulations described in
Table~\ref{tablesimulationsdel}. Note that for all plots in this paper, we normalize the divergence ($\theta=\nabla\cdot\uv$) and the vorticity ($\wv=\nabla\times\uv$) so that they refer to the dimensionless quantities, i.e. 
\beq
\nabla\cdot\uv/{\cal H}f \ \ \ \ \ {\rm and} \ \ \ \ \ 
\nabla\times\uv/{\cal H}f,
\label{norm}
\eeq
where ${\cal H} = d\ln a/d\tau$ and $f=d\ln D_+/d \ln a$ is the logarithmic derivative of the linear growth factor $D_+$ with respect to the scale factor $a$. This change of units is convenient since in linear theory, the divergence normalized in this way equals minus the dimensionless overdensity, i.e.  $\theta=-{\cal H}f \delta$, with $\delta=\delta\rho/ \bar{\rho}$.

Figure~\ref{rptvsdel} shows the average, over the 50 realizations of run LR1280, of the divergence power spectrum at $z=0$ compared with the 2-loop RPT prediction. The divergence power spectrum behaves as expected theoretically, with suppressed growth compared to linear theory, although one can notice significant deviations from RPT, which will be explored in detail elsewhere. Note that the non-linear effects in the power spectrum are observable on scales with $k\gtrsim 0.01 \Mpc$, unlike the case for the density field. This is expected for two reasons: first, the velocity divergence propagator decays faster than for the density field, damping the linear spectrum faster with $k$~\cite{2006PhRvD..73f3520C}; second, the mode coupling power generated at small scales is smaller than for the density field, avoiding the accidental cancellation of nonlinear effects present in the density power spectrum~\cite{2008PhRvD..77b3533C}. This qualitative behavior is also predicted by standard PT~\cite{2004PhRvD..70h3007S,2008arXiv0806.0971P}. Clearly, as discussed in~\cite{2004PhRvD..70h3007S}, assuming that density and velocity divergence power spectra are equal (as often done for redshift distortions) is not a very good approximation, even at large scales.

Figure~\ref{vortilus} illustrates the spatial distribution of  velocity and density estimations from the HR160 run at $z=0$. The panels show overdensity, divergence and vorticity on a $1 \Mpc$-thick cross-section of the simulation. Also, the overdensity corresponding to {\em halo particles} (particles inside dark matter halos) is shown.  Even though the overdensity field can take on values up to a few hundreds, its scale was chosen to go up to $\delta=3$ because the dark matter halos are small compared to the scale of this figure and increasing the upper scale limit would just hide the lower density structures. 

We can see that the divergence field is, not surprisingly, remarkably similar to the density field. However, the structures in the velocity divergence have in general lower amplitude and are more extended in space, as expected from the power spectrum results discussed above. It is interesting to note that, at the halo positions, the divergence tends to be smaller than in the still collapsing regions, as it should be. On the other hand, the vorticity field fluctuates in sign on scales of the order of $\sim 1 \Mpc$ (roughly as expected from theoretical arguments, see Fig.~8 in~\cite{1999A&A...343..663P}), and it is concentrated on collapsing regions, where shell-crossing is currently occurring. There are no large-scale coherent fluctuations in vorticity, so we expect the vorticity power spectrum to be much smaller than the divergence power on large scales, as we now discuss.

\begin{table*}
\begin{center}
\begin{tabular}{l@{\hspace{1cm}}c@{\hspace{1cm}}c@{\hspace{1cm}}c@
{\hspace{1cm}}c@{\hspace{1cm}}l}\hline
Name & $L_{\rm box}$ & $N_{part}$ & $m_{\rm par}$&
$N_{\rm realizations}$&softening\\\hline
LR1280 & 1280 & $640^3$ & 59.94 & 50 & 0.07\\
LR512 & 512 & $256^3$ & 59.94 & 1 & 0.2\\
MR512 & 512 & $512^3$ & 7.49 & 1 & 0.04\\
MR320 & 320 & $640^3$ & 0.94 & 1 & 0.015\\
HR160 & 160 & $640^3$ & 0.12 & 1 & 0.00625\\
SHR160 & 160 & $1024^3$ & 0.029 & 1 & 0.00625\\\hline
\end{tabular}
\end{center}
\caption{All our simulations have $\Omega_m=0.27$, $\Omega_\Lambda=0.73$, $\Omega_b=0.046$, $h=0.72$ and  $\sigma_8(z=0)=0.9$. They were run using the {\sf Gadget2} code~\cite{2005MNRAS.364.1105S}. $L_{box}$ is in units of $\Mpc$ and $m_{par}$ is the particle mass in units of $10^{10}h^{-1}M_\odot$. }
\label{tablesimulationsdel}
\end{table*}

\subsection{Dependence  on Mass Resolution}
\begin{figure}
\begin{center}
\includegraphics[width=75mm]{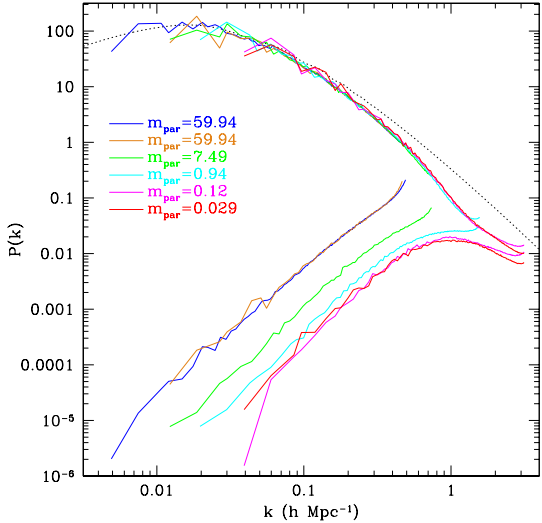}
\end{center}
\caption{Dependence on mass resolution of the velocity divergence and vorticity power spectra. Particle masses are labeled in units of $10^{10}h^{-1}M_\odot$, see Table~\ref{tablesimulationsdel} for more details on the simulations. While the divergence power spectrum does not depend on mass resolution, the vorticity power spectrum does show  significant sensitivity. However, as mass resolution increases it converges (when $m_{\rm par}$ is below $\sim10^{9}h^{-1}M_\odot$) to a stable answer.}
\label{res_dependence}
\end{figure}

Figure~\ref{res_dependence} shows the power spectrum of divergence and vorticity obtained from the Delaunay method from different simulations (see  Table~\ref{tablesimulationsdel}). The velocity field is dominated, especially on large scales, by its irrotational component, consistent with the spatial distribution seen in Fig.~\ref{vortilus}. We see from Fig.~\ref{res_dependence} that the divergence power spectra measured over a broad range of volume, number of particles and mass resolution simulations match consistently. 

The estimate of the vorticity power spectrum, on the other hand, appears not to be so robust: it shows a clear monotonic dependence on the mass resolution. We verified that this dependence was not an artifact of the Delaunay method by comparing these results to the ones obtained from the CIC mass-weighted scheme. We observed that these two methods agree on the mass resolution dependence of the vorticity power spectrum (not shown in Fig.~\ref{res_dependence} for clarity). Thus we believe the dependence on mass resolution of the measured vorticity is real and may be due to insufficient sampling of collapsing regions~\footnote{As the writing of this paper was being completed, the work~\cite{2008arXiv0808.0003P} was submitted, in which they seem to find a similar dependence of the vorticity power on mass resolution.}. However, as the particle mass goes below $m_{\rm par} \sim 10^{9} h^{-1}M_\odot$, the vorticity power spectrum eventually converges.

Also, we check for aliasing effects, discussed in detail in the Appendix. Our estimates for the spurious aliased vorticity based on Eq.~(\ref{eqaliasing}) are at least two orders of magnitude lower than the measured vorticity from the simulations. Although this shows the spurious vorticity it is not a sampling issue in the {\em measurement} of the power spectrum, aliasing may be an issue in the low resolution simulations (which have a coarser PM grid) {\em during} time evolution, since the spurious power is close to the expected $k^2$ behavior (see section~\ref{secaliasing} in the Appendix) at all scales.  From here on, our results will be based only on the higher mass resolution runs. 

Regarding the dependence on resolution of the divergence power spectrum, close inspection of Fig.~\ref{res_dependence} seems to indicate that the higher mass resolution runs have a larger power spectrum than lower mass resolution runs by about $5-10\%$. However, one must keep in mind the higher mass resolution runs have substantially smaller box sizes (see  Table~\ref{tablesimulationsdel}). For box sizes smaller than about $\sim 300 \Mpc$, RPT predicts that the propagator is seriously affected by the finite volume of the simulation (see Fig.~6 in~\cite{2006PhRvD..73f3520C}). For such boxes, the damping of the linear spectrum by the propagator is much less severe, and while the mode coupling power is somewhat larger, it cannot compete with the nearly exponential scale dependence of the propagator. Thus one expects to see  higher divergence power in smaller boxes. This is confirmed further by looking at simulations of same box size (LR512 and MR512, on one hand, and HR160 and SHR160 on the other). The ratio of the power spectra in boxes of the same size but very different mass resolutions does not show any significant (percent-level) deviation from unity. 

Finally, note that finite volume effects are not expected to affect the vorticity power spectrum, as it is dominated by small scale structures. That the vorticity is sensitive to mass resolution rather than box size is clear from comparing the LR512 and MR512 results, which differ by a factor of 8 in mass resolution but have the same box size. Figure~\ref{res_dependence} shows their vorticity power spectra differ by a factor of about four.

It is worth noting that the velocity power spectrum obeys $P_v(k)=k^2 [P_\theta(k)+P_w(k)]$.  Thus, the resolution dependence of $P_w$ seen in Fig.~\ref{res_dependence} means that when $P_w$ is comparable to $P_\theta$, a similar dependence  on resolution affects the velocity power spectrum. Therefore, spurious vorticity can lead to an overestimate of the velocity power spectrum at nonlinear scales.

\subsection{Time Dependence}
\label{sec:a_dependence}
\begin{figure}
\begin{center}
\includegraphics[width=75mm]{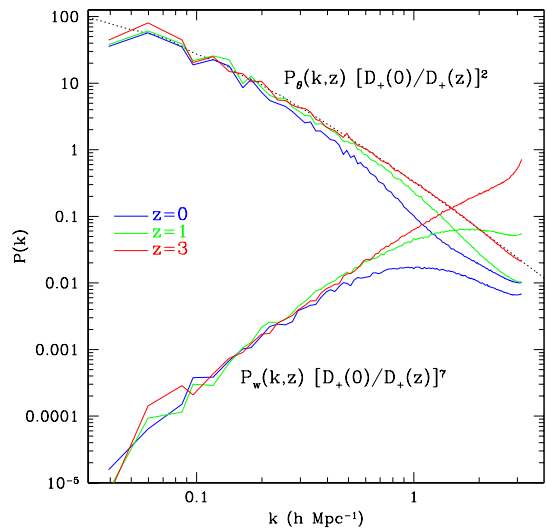}
\end{center}
\caption{Time dependence of the divergence and vorticity power spectra. The divergence power spectrum at $z=1$ and $z=3$ are linearly extrapolated to $z=0$ for comparison. The vorticity power spectrum was similarly scaled using Eq.~(\protect\ref{vort_a_dependence}) with $n_w=7$. In the non-linear regime, both divergence and vorticity grow slower than the large-scale  extrapolation.} 
\label{a_dependence} 
\end{figure}

As will be shown later, in order to calculate how much vorticity affects the evolution of the density power spectrum, it is necessary to determine the time dependence of the vorticity power spectrum. In linear theory, the divergence power spectrum evolves according to
\beq
P_\theta(k,z) \simeq [D_+(z)]^2\, P_{\theta 0}(k),
\label{div_a_dependence}
\eeq
where $P_{\theta 0}(k)$ is the initial (post-recombination) divergence power spectrum and $D_+(z)$ is the linear growth factor measured away from the initial conditions. For the vorticity power spectrum, we propose that in the large-scale limit
\beq
P_w(k,z) \propto [D_+(z)]^{n_w}.
\label{vort_a_dependence}
\eeq

From the vorticity power spectra estimates at $z=0,1,3$, we find that the best fit for Eq.~(\ref{vort_a_dependence}) is given by $n_w=7 \pm 0.3$. Figure~\ref{a_dependence} shows the divergence power spectrum at $z=0,1,3$ extrapolated to $z=0$ using Eq.~(\ref{div_a_dependence}), and the vorticity power spectrum extrapolated using Eq.~(\ref{vort_a_dependence}) with $n_w=7$. It can be seen that at large scales the extrapolated outputs agree very well while in the non-linear regime these simple scalings break down, as expected, with the growth slowing down compared to large scales. In the case of the divergence power spectrum, this behavior can be understood reasonably well from RPT, see Fig.~\ref{rptvsdel}.  A detailed discussion of velocity statistics and RPT will be presented elsewhere.

For the vorticity, little is known from first principles. The exception is the work in~\cite{1999A&A...343..663P}, were the {\em rms} vorticity is calculated for CDM spectra during first orbit crossing using the Zel'dovich approximation (see also~\cite{2002MNRAS.334..435D} for a calculation of {\em rms} vorticty for scale-free models). They found their estimates were much smaller than found in simulations, which given their mass resolution at the time is not surprising (see Fig.~\ref{res_dependence}). However, one can {\em very roughly} estimate the vorticity power spectrum at large scales generated by orbit crossing by postulating that orbit mixing creates in such regions a velocity field that is approximately the result of mass weighting the single stream velocities (see discussion related to Eq.~\ref{CmultiS} below). Then we expect $\wv \simeq (1+\delta)^{-1} f_v \nabla \times [(1+\delta)\vv]$, where $f_v$ is only nonzero in regions where orbit crossing occurs, and on average can be thought as the fraction of the volume that undergoes orbit crossing, an increasing function of time. Then the vorticity power spectrum reads~\cite{2001NYASA.927...13S}

\beqa
P_w(k,z) &\sim & [f_v(z)]^2 \int d^3q\frac{(\kv \times \qv)^2}{q^4} \Big[ P_\delta(|\kv-\qv|)P_\theta(q) \nonumber \\
& -& \frac{q^2}{(\kv-\qv)^2} P_{\rm x}(|\kv-\qv|)P_{\rm x}(q) \Big],
\label{Pwguess}
\eeqa
where $P_{\rm x}$ is the cross spectrum between $\delta$ and $\theta$. In the low-$k$ limit this reduces to

\beqa
P_w(k,z) &\sim & [f_v(z)]^2 \int d^3q\frac{(\kv \times \qv)^2}{q^4} \Big[ P_\delta(q)P_\theta(q) \nonumber \\
& -&  P_{\rm x}(q)P_{\rm x}(q) \Big] \propto  k^2\, [f_v(z)]^2\, [D_+(z)]^6, \nonumber \\ 
\label{Pwguess2}
\eeqa
where in the last step we have assumed the velocities are normalized as in Eq.~(\ref{norm}) and used that the square brackets vanish in linear theory, so the leading nonzero contribution comes from one-loop PT (which induces a $D_+^4$ time dependence in the power spectra beyond leading order). Despite the crude approximations made in arriving to Eq.~(\ref{Pwguess2}), the scale and time dependence of the large-scale vorticity power spectrum seen in Fig.~\ref{a_dependence} may be explained qualitatively along these lines.

\subsection{Impact of Virial Velocities}
\begin{figure}
\begin{center}
\includegraphics[width=75mm]{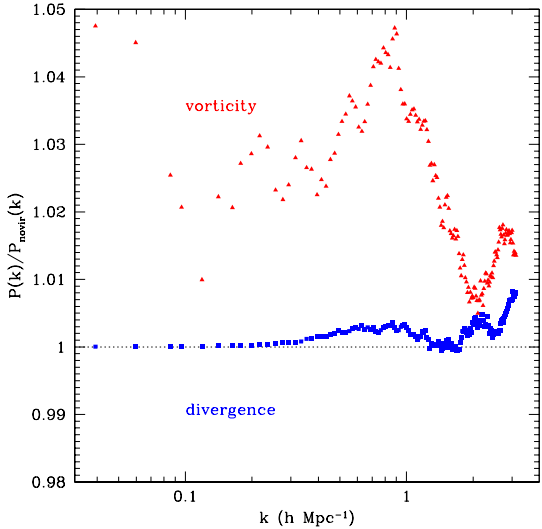}
\end{center}
\caption{Comparison between divergence and vorticity power spectra of the simulation to the power spectra obtained by replacing velocities of particles inside halos by the center of mass velocity of the parent halo, thus setting virial velocities to zero. We can see that the divergence is mostly unaffected, while the vorticity differences are less than $5\%$.}
\label{novirtest}
\end{figure}

In Figure~\ref{vortilus}, it can be observed that the velocity field is rotational in high-density collapsing regions. Compare, for instance, the lower panels against the top right panel for which only particles belonging to halos are shown. On the upper left corner of the simulation box, there is a large filamentary structure. We can see that the vorticity occurs mainly on the outskirts of virialized objects. This suggests that the fraction of the vorticity power spectrum coming from virialized regions themselves is not very big. To check this, we took the HR160 simulation and replaced the particle velocities belonging to halos by the center-of-mass velocity of the corresponding halo, thus eliminating the velocity dispersion  of all halos. We measured divergence and vorticity power spectra and compared them to those of the unmodified HR160 simulation. The results are shown in Figure~\ref{novirtest}. It can be seen that at the scales we probe the divergence power spectrum is essentially not affected by the virial velocities, and the vorticity power spectrum is reduced by less than 5\%. It is important though to keep in mind that our measurements on the HR160 simulations are done in a grid of size 160, so the contribution from scales less than $\sim 1 \Mpc$ are not included; this roughly corresponds to ignoring halos of $m \la 10^{14} h^{-1} M_\odot$.

That the vorticity power spectrum is not very sensitive to virial velocities may be understood by considering the vorticity evolution equation (\cite{2002PhR...367....1B}, see also Eq.~\ref{eqn:vortconservation_2} below)
\beq
\frac{\partial\wv}{\partial\tau}+{\cal H}\wv -
\nabla\times\left[\uv\times\wv\right] =
\nabla\times\left(\frac{1}{\rho}\nabla\cdot(\rho\,\vec\sigma)\right),
\label{vort_eq_ex}
\eeq
where $\rho \sigma_{ij}$ is the stress tensor induced by orbit crossing. In a virialized object, where the phase-space distribution function is approximately Maxwellian with velocity dispersion related to halo mass through $\sigma^2_{vir}\sim Gm/r_{vir}$ ($r_{vir}\propto m^{1/3}$), we expect the stress tensor to be reasonably  well approximated by an equation of state of the form $\rho\sigma_{ij}\sim -p\delta_{ij}$, where $p$ is a density-dependent pressure; in fact, for an isothermal sphere $p\propto \rho \sigma^2_{vir}$ with $\sigma^2_{vir}$ independent of spatial position. In that case, the forcing term of Eq.~(\ref{vort_eq_ex}) can be written as 
\beq
\nabla\times\left(\frac{1}{\rho}\nabla\cdot(\rho\,\vec\sigma)\right)=
\frac{\nabla\rho}{\rho^2}\times\nabla p \approx 0,
\eeq
where the last step follows from the density-dependence of the pressure. Therefore, although these approximations are not totally realistic in practice, they may help explain small vorticity sourcing from virialized dark matter halos at the scales we probe.

\section{Generation of Vorticity and Velocity Dispersion}
\label{orbitcross}

\subsection{Beyond PPF}

One of the main goals of this paper is to estimate the corrections to the pressureless perfect fluid (PPF) predictions for power spectra of density and velocity fields at large scales due to orbit crossing at small scales. In order to do so, we have to estimate the corrections to the equations of motion beyond PPF that result from solving the Vlasov equation for the phase-space distribution function (hereafter DF) $f(\xv,\pv,\tau)$,

\beq
\frac{\partial f}{\partial \tau}+\frac{\pv}{a}\cdot \nabla f - a \nabla \phi \cdot \frac{\partial f}{\partial \pv}=0,
\label{Vlasov}
\eeq
where $\pv$ is the momentum per unit mass and $\phi$ the gravitational potential. Equation~(\ref{Vlasov}) says that the DF is conserved ($df/d\tau=0$) along its characteristics,
\beq
\frac{d \xv}{d\tau} = \frac{\pv}{a},\ \ \ \ \ 
\frac{d \pv}{d\tau}=-a \nabla \phi,
\label{Hamilton}
\eeq
which are the Hamilton equations of motion, that can be combined to give the familiar result,
\beq
\frac{d^2\xv}{d\tau^2}+{\cal H} \frac{d\xv}{d\tau} = - \nabla \phi.
\label{partEOM}
\eeq

Cosmological N-body simulations solve the Vlasov equation by discretizing the DF using particles that follow the characteristics, Eqs.~(\ref{Hamilton}) or (\ref{partEOM}). To make connection with the PPF equations of motion, one may take moments (or, more precisely, cumulants) of the Vlasov equation. The first few cumulants of the DF are the (comoving) density field,
\beq
(1+\delta) = \int f(\pv) \ d^3p,
\label{deDF}
\eeq
where to simplify notation we avoid displaying the space and time arguments everywhere; the velocity field $\uv$,
\beq
(1+\delta)\, \uv = \int f(\pv) \, \frac{\pv}{a} \ d^3p,
\label{velDF}
\eeq
and the stress tensor $T_{ij}\equiv (1+\delta)\sigma_{ij}$,
\beq
(1+\delta)\, \sigma_{ij} = \int f(\pv)\, \frac{p_i p_j}{a^2}\, d^3p -(1+\delta)\, u_i u_j,
\label{stressDF}
\eeq
where the velocity dispersion tensor $\sigma_{ij}$ describes isotropic and anisotropic velocity dispersion. Before we derive equations of motions for these quantities, it is useful to introduce the cumulant generating function, which generates all these objects. As it is usual in statistics of large-scale structure (see e.g.~\cite{2002PhR...367....1B}), the cumulant generating function ${\cal C}$ is given in terms of the moment generating function ${\cal M}$ by
\beq
{\cal C}(\lv)=\ln {\cal M}(\lv), \ \ \ \ \ 
{\cal M}(\lv) \equiv \int {\rm e}^{\lv\cdot \pv/a} \ f(\pv)\, d^3p,
\label{Gen}
\eeq
where moments are obtained by successive derivatives of ${\cal M}$ with respect to the external parameter $\lv$,
\beq
(\nabla_{l_{i_1}} \ldots \nabla_{l_{i_n}} {\cal M})_0 = (1+\delta)\  m^{(n)}_{i_1\ldots i_n}
\label{moments}
\eeq
where $(\ldots)_0$  means evaluating quantities at $\lv=0$, and  $m^{(0)}=1$, ${\cal M}_0 =(1+\delta)$, $m^{(1)}_i=u_i$, $m^{(2)}_{ij}=u_iu_j+\sigma_{ij}$. Cumulants are statistically independent objects at each order and can be obtained similarly by differentiation of ${\cal C}$,  
\beq
(\nabla_{l_{i_1}} \ldots \nabla_{l_{i_n}} {\cal C})_0 = c^{(n)}_{i_1\ldots i_n},
\label{cumulants}
\eeq
with  $c^{(0)}={\cal C}_0 =\ln(1+\delta)$, $c^{(1)}_i=u_i$, $c^{(2)}_{ij}=\sigma_{ij}$.

From the Vlasov equation, Eq.~(\ref{Vlasov}), and Eq.~(\ref{Gen}) it is straightforward to derive equations of motion for the generating functions. For ${\cal C}$ we have,

\beqa
\frac{\partial {\cal C}}{\partial \tau}+{\cal H}\,( \lv \cdot \nabla_{\lv}){\cal C}+\nabla{\cal C}\cdot \nabla_{\lv}{\cal C}+(\nabla\cdot\nabla_\lv){\cal C}
&=&-\lv\cdot \nabla\phi. \nonumber \\ & & 
\label{EOMcumgen}
\eeqa
This is a nonlinear partial differential equation for ${\cal C}(\xv,\tau,\lv)$; however, all we are interested in is what happens in the neighborhood of $\lv=0$, i.e. the derivatives of ${\cal C}$ at $\lv=0$, see Eq.~(\ref{cumulants}).

The equations of motion beyond the PPF approximation readily follow from Eq.~(\ref{EOMcumgen}). Setting $\lv=0$ we obtain the continuity equation,

\beq
\frac{\partial \delta}{\partial \tau}+\nabla\cdot [(1+\delta) \uv]=0,
\label{continuity}
\eeq
whereas taking the first derivative we obtain momentum conservation,

\beq
\frac{\partial u_i}{\partial \tau} + {\cal H} u_i +( \uv \cdot \nabla) u_i = -\nabla \phi - {1\over \rho} \nabla_j (\rho \sigma_{ij}),
\label{momentum}
\eeq
where $\rho \equiv 1+\delta$, while the evolution of the velocity dispersion tensor is obtained from  Eq.~(\ref{EOMcumgen}) by applying second derivatives,

\bea
\label{stressEOM}
\frac{\partial\sigma_{ij}}{\partial\tau}&+&2{\cal  H}\sigma_{ij}+
(\uv\cdot\nabla)\sigma_{ij}\\
&+&\sigma_{jk} \nabla_k
u_i+\sigma_{ik} \nabla_k u_j = -\frac{1}{\rho} \nabla_k(\rho \Pi_{ijk}),
\nonumber
\eea
where $\Pi_{ijk}\equiv c^{(3)}_{ijk}$ is the third cumulant of the DF, see Eq.~(\ref{cumulants}). By applying successive derivatives with respect to $\lv$ in Eq.~(\ref{EOMcumgen}) one thus generates an infinite hierarchy of equations of motion for the cumulants of the DF (hereafter cumulant hierarchy). The hierarchy is infinite because at finite order is never closed, the cumulant of order $n$ depends on that of order $n+1$.

\subsection{The Cumulant Hierarchy and Orbit Crossing}
\label{hierarchical}

Such an infinite hierarchy is very difficult to solve. The PPF approximation truncates the hierarchy assuming that the second and higher order cumulants of the DF are zero, thus $\sigma_{ij}=0$, $\Pi_{ijk}=0$ in Eqs.~(\ref{momentum}-\ref{stressEOM}) and so on. This is equivalent to assuming the DF takes the form,
\beq
f(\xv,\pv,\tau)=[1+\delta(\xv,\tau)] \, \delta_{\rm D}[\pv-a\, \uv(\xv,\tau)],
\label{DF1stream}
\eeq
for which ${\cal C}(\lv)=\ln(1+\delta)+\lv\cdot \uv$, and clearly all cumulants of order larger than one vanish. Note that the PPF approximation appears to be {\em self-consistent}, i.e. assuming that $\sigma_{ij}$ and higher order cumulants vanish at a given time is preserved by the hierarchy. This is so because there are no linear or nonlinear terms in the equations of motion for such cumulants that solely involve the density and/or velocity fields as sources. This can be readily seen from the structure of Eq.~(\ref{EOMcumgen}), after operating with two or more derivatives $\nabla_{\lv}$. In other words, if ${\cal C}$ initially only contains linear terms in $\lv$, Eq.~(\ref{EOMcumgen}) will not generate higher powers in $\lv$.

This situation is, however, unstable under perturbations. If somehow ${\cal C}$ develops a quadratic contribution in $\lv$, then the nonlinear term in Eq.~(\ref{EOMcumgen}) generates a cubic term, and this in turn generates higher orders, and so on. Therefore, once velocity dispersion ``turns on", all higher order cumulants do so as well. Thus a priori it is not a self-consistent truncation to include a non-zero $\sigma_{ij}$ and ignore $\Pi_{ijk}$ (which {\em is} sourced by terms solely dependent on $\sigma_{ij}$) and higher-order cumulants. This truncation may, however, become a good approximation in some situations, e.g. at large scales.

Physically it is  expected that even for perfectly ``cold" initial conditions where the DF is given initially by Eq.~(\ref{DF1stream}), orbit crossing during time evolution will generate a nontrivial stress tensor and higher-order cumulants, while as discussed above the cumulant hierarchy does not seem to allow for this. Given that such a result from the cumulant hierarchy is unstable to small perturbations away from cold initial conditions, any subtlety in going from the Vlasov equation to Eq.~(\ref{EOMcumgen}) may alter the conclusions. A more careful look at orbit crossing in this context shows that this suspicion is well founded.

To see how orbit crossing generates a nontrivial DF from cold initial conditions, consider the formal solution of the Vlasov equation expressing the conservation of the DF along the characteristics,

\beq
f(\xv,\pv,\tau) = f_0(\Xv_0,\Pv_0),
\label{VlasovSol}
\eeq
where $f_0$ is the initial DF, and,
\beq
\Xv_0\equiv \Xv_0(\xv,\pv,\tau), \ \ \ \ \ 
\Pv_0\equiv \Pv_0(\xv,\pv,\tau)
\label{HamMap}
\eeq
are the initial positions and momenta which when evolved by the equations of motion until time $\tau$ (Eqs.~\ref{Hamilton}) lead to $\xv$ and $\pv$. That is, time evolution maps $(\Xv_0,\Pv_0)$ to $(\xv,\pv)$ at time $\tau$, and this mapping is invertible because in {\em phase space} trajectories never intersect for a Hamiltonian flow.

\begin{figure}[t!]
\begin{center}
\includegraphics[width=75mm]{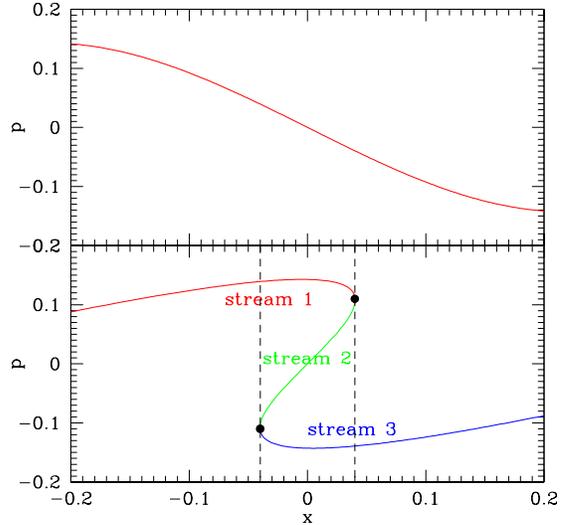}
\end{center}
\caption{Phase space sketch of generation of multiple streams due to orbit crossing. The top panel shows the zero width DF after evolution from cold initial conditions before orbit crossing. The bottom panel shows the DF after orbit crossing (which occurs at the moment when ``stream 2" is infinitesimal and perpendicular to the x-axis, while the two vertical lines coincide, as well as the filled circles). In between the two vertical lines there are three ``streams": that's the region of space where multistreaming is present, and e.g. velocity dispersion is generated. The intersections of the vertical lines with the streams at the filled circles denote where the derivative of the curve is infinite and thus at such positions the density field (projection of the DF onto the x-axis)  is singular.} 
\label{Caustics}
\end{figure}

However, orbits clearly can (and do) cross in {\em configuration space}, i.e. at time $\tau$ and position $\xv$ there may be more than one orbit (with different $\pv$'s) that trace back to different initial conditions $(\Xv_0,\Pv_0)$. If we start from cold initial conditions, $f_0$ satisfies Eq.~(\ref{DF1stream}), and after time evolution the DF reads, from Eq.~(\ref{VlasovSol})

\beq
f(\xv,\pv,\tau) =[1+\delta_0(\Xv_0)]\, \delta_{\rm D}[\Pv_0-\uv_0(\Xv_0)],
\label{fsolColdICs}
\eeq
where we have set $a_0\equiv 1$ and $\delta_0$ and $\uv_0$ are the initial density and velocity fields obtained, typically, from Gaussian random field initial conditions. Now we are ready to see the effect of orbit crossing on the cumulant generating function $C(\lv)$, Eq.~(\ref{Gen}). As long as orbits do not cross, since $f_0$ has zero width then at fixed $(\xv,\tau)$ there is a unique $\pv$ that contributes to the momentum integral (see top panel in Fig.~\ref{Caustics}). Thus the argument of the delta function in Eq.~(\ref{fsolColdICs}) can be linearly related to $\pv$, and ${\cal C}$ preserves its linear dependence on $\lv$ and no higher-order cumulants are generated. 

However, as soon as orbits cross, there are many $\pv$'s  at fixed $(\xv,\tau)$ and thus many roots of the argument of the delta function in Eq.~(\ref{fsolColdICs}), each of them corresponding to one ``stream", see bottom panel in Fig.~\ref{Caustics}. As a result, the cumulant generating function reads instead

\beq
{\cal C}(\xv,\tau,\lv) = \ln \Big[ \sum_{\rm streams~at~\xv} (1+\delta_s) \ {\rm e}^{\lv \cdot \uv_s} \Big],
\label{CmultiS}
\eeq
where we have written schematically $\delta_s$ and $\uv_s$ for the density and velocity fields of each stream, which can be obtained by projecting each piece of the DF separately, see Fig.~\ref{Caustics}. Clearly, if the number of streams is larger than one, ${\cal C}$ is a fully nonlinear function of $\lv$ and all cumulants have been generated simultaneously by orbit crossing. Note that the number of streams at position $\xv$ and time $\tau$ is a random field that depends on initial conditions and cosmological parameters, see e.g.~\cite{1994ApJ...420...44K} for a calculation of the mean number of streams from Gaussian initial conditions.

We now see there is an apparent contradiction between the Vlasov formal solution and the time evolution of the cumulant hierarchy. The root of this can be traced back to the well known fact that at orbit crossing the density field becomes singular~\footnote{For example, in the Zel'dovich approximation it can be shown that the density field diverges as $|\xv-\xv_c|^{-1/2}$ inside a multi-streaming region, where $\xv_c$ corresponds to the position of the  caustic formation. For example, all stream densities $\delta_s$ in Eq.~(\ref{CmultiS}) will have such singularities, in addition the velocity field is multivalued. See e.g.~\cite{1989RvMP...61..185S} for a review.}, therefore the cumulant hierarchy must be supplemented with a regularization procedure in order to follow time evolution after orbit crossing.  This must restore the agreement with the formal solution of the Vlasov equation, which does not suffer from development of singularities (which invalidate the hierarchy because projection in momenta does not commute with time evolution).

Does this matter in practice? After all CDM has a very small but non-zero velocity dispersion, which will automatically regularize the singularities in the cumulant hierarchy that arise at orbit crossing. However, it seems unlikely  that  the final answer for {\em large-scale} density and velocity fields after orbit crossing should depend sensitively on the velocity dispersion of the CDM particles (if this were so CDM N-body simulations would be incorrect); although of course such effects are important for warm dark matter candidates. Rather, it should be the self-gravity of regions that undergo orbit crossing that leads self-consistently to velocity dispersion and higher-order cumulants that regularizes the singularities. 

To carry out such self-consistent regularization, one can proceed in at least two ways: 1) introduce some non-zero initial velocity dispersion $\sigma_0$ and use the hierarchy, which does not develop singularities in this case, to evolve the system forward in time. To make predictions for systems with negligible initial velocity dispersion such as CDM one must take into account that the  $\sigma_0\rightarrow 0^+$ limit is nontrivial and one should get finite corrections for infinitesimal $\sigma_0$. 2) since mass does not diverge in caustics, one can work with coarse grained variables (smoothed over some small scale). To find the equations of motion for smoothed quantities one must take into account  that coarse graining does not commute with time evolution, and the coarse graining scale must be picked so that physics at large scales is invariant with respect to the degrees of freedom that are integrated over at small scales. See~\cite{2000PhRvD..62j3501D} for some steps in this direction.

In either case, the net result of regularization is that higher-order cumulants of the DF {\em will} be sourced by density and velocity fields, which leads to an effective equation of state for dark matter. On the other hand, one would still have to implement a consistent closure of the hierarchy. 

In this paper we proceed in a different fashion, by measuring the stress tensor directly from numerical simulations. We then close the hierarchy by using the  measured stress tensor in the momentum conservation equation, Eq.~(\ref{momentum}). We  start our measurements at $z=3$ and assume that the dark matter has undergone sufficient shell crossing before $z=3$ that future caustics are not singular, and thus Eqs.~(\ref{continuity}-\ref{momentum}) are valid. To extrapolate backwards in time we use the time dependence found from $z=3$ to $z=0$ and assume it is valid earlier down to the initial conditions. Since we are interested in the large-scale statistics of density and velocity fields, the deviations from the PPF approximation are very small before $z=3$, thus this should be a reasonable approach. Before we describe how we implement such a procedure in detail, we must explain how we estimate the stress tensor ($T_{ij}=\rho \sigma_{ij}$) from the numerical simulations.

\subsection{Estimating The Stress Tensor}
\label{secveldisp}

In the Appendix~\ref{secdelaunay}, we discuss how the Delaunay method is a reliable algorithm for recovering the velocity field from numerical simulations. However, as we will now argue, it is not adequate for estimating the velocity dispersion tensor $\sigma_{ij}$. The Delaunay method is optimized for interpolating the velocity field on arbitrary points in the simulation. This procedure recovers a continuous field that is to be interpreted as the mean velocity field. Then, if one tried to estimate the dispersion in a volume $\Delta V$ with an estimator of the
form 
\bea
\sigma_{\rm Del}&=&\frac{1}{\Delta V}\int d^3x\ 
\uv_{\rm Del}(\xv)\otimes\uv_{\rm Del}(\xv) - \nonumber \\
&&\bar\uv_{\rm Del}\otimes\bar\uv_{\rm Del}
\eea
where $\uv_{\rm Del}(\xv)$ is the Delaunay-interpolated velocity field and $\bar\uv_{\rm Del}$ is the average on the volume $\Delta V$ of $\uv_{\rm Del}(\xv)$, the integral would be dominated by the Delaunay linear approximation to the mean velocity, contaminating the true velocity dispersion coming from multi-streaming. 

Our strategy to estimate the velocity dispersion is based on the fact  that particles in numerical simulations are a sample of the phase-space distribution function $f(\xv,\pv,\tau)$ (see Eq.~\ref{Vlasov}). Consider, for instance, a small volume $\Delta V$ in which the distribution function is nearly constant ($\nabla f\approx 0$). Our ansatz is that the simulation particles in $\Delta V$ are sampled from the probability distribution given $\bar{f}_{\Delta V}(\pv,\tau)$ by
\beq
\bar{f}_{\Delta V}(\pv,\tau)=\frac{1}{\rho\Delta V}\int_{\Delta V}
d^3x\ f(\xv,\pv,\tau),
\label{reducedf}
\eeq
where $\rho$ is the density in the small volume (which is constant due to the ansatz). Therefore, the volume-averaged velocity dispersion tensor $\bar{\sigma}_{ij}$ in that small volume can be written as
\bea
\bar\sigma_{ij}&=&\frac{1}{\Delta V}
                      \int_{\Delta V}d^3x\ \sigma_{ij}(\xv)\nonumber\\
&=&\int d^3p \frac{p_i p_j}{a^2}\left[\frac{1}{\rho\Delta V}
   \int_{\Delta V}d^3x\,f(\xv,\pv,\tau)\right]-\nonumber\\
&&\bar u_i \bar u_j,
\eea
where $\bar u_i$ and $\bar u_j$ are the mean velocity field in the volume $\Delta V$ (the velocity field is also constant on $\Delta V$ due to the ansatz). The term in square brackets is the one defined in Eq.~(\ref{reducedf}). Since we assume the particles in $\Delta V$ are sampled from that distribution, we can write
\beq
\bar\sigma_{ij}=\frac{1}{N}\sum_{n=1}^N
                 u^{(n)}_i u^{(n)}_j - \bar u_i \bar u_j,
\label{sigmaestimator}
\eeq
where the sum is over the $N$ particles in the small volume $\Delta V$ and $\uv^{(n)}$ is the velocity of the n-th particle.

Equation~(\ref{sigmaestimator}) will serve us as an estimator for the velocity dispersion on a small region of constant phase-space distribution function. To obtain the volume-weighted dispersion on a larger volume, we simply break that volume into regions where Eq.~(\ref{sigmaestimator})  is valid and then volume-average the results. For a different approach to estimating velocity dispersion from simulations see \cite{2003AN....324..560D,2004A&A...419..425D,2006MNRAS.371.1959K}. 

In practice, we want to compute the volume-averaged velocity dispersion on a grid, similarly to Eq.~(\ref{delaunay_kernel}) for the mean velocity. We use the following recursive algorithm to find that average in a given cell: 

\begin{itemize}
\item[i)] First determine whether cell is homogeneous enough. To do this compute density $\rho_i$ in each octant of the current cell. Then compute density standard deviation, $s=\sqrt{\frac{1}{8}\sum_{i=1}^8(\rho_i-\bar\rho)^2}$. 
\item[ii)] If $s<s_{\rm threshold}$ or $N<N_{\rm partmin}$, we can use Eq.~(\ref{sigmaestimator}). Then, compute $\bar\uv=\frac{1}{N}\sum_{n=1}^N \uv^{(n)}_i$ and $\bar\sigma=\frac{1}{N}\sum_{n=1}^N\uv^{(n)}\otimes\uv^{(n)} -  \bar\uv\otimes\bar\uv$.
\item[iii)] Else, the cell is not homogeneous. Following this three-step algorithm, calculate recursively the velocity dispersion $\bar\sigma^{(i)}$ of each octant, then $\bar\sigma=\frac{1}{8}\sum_{i=1}^8\bar\sigma^{(i)}$.
\end{itemize}
Here, $s_{\rm threshold}$ is the minimum standard deviation of the octants density that defines the criterion for homogeneity (we use typically values of 30\%-50\% of the octants mean density). The purpose of $N_{\rm partmin}$ is to avoid breaking cells with too few particles into octants (we use $N_{\rm partmin}\simeq 5$). Note that this algorithm is adaptive: it resolves the inhomogeneous dense regions to find the subregions where Eq.~(\ref{sigmaestimator}) holds. The method has two free parameters, which varied on a reasonable range produce shifts in the observed velocity dispersion tensor up to 40\% (this will suffice for our purposes, as we shall see below). This is understandable, since these parameters compensate for the finite resolution (and sampling) of the phase-space distribution function. Another drawback of the method is that it suffers from Poisson noise in low density regions. However, on large scales, since we are averaging over many cells, we expect these errors to be rather small. 

The stress tensor estimated in this way satisfies a nontrivial sanity check, as we will show below. Its vector modes source the growth of vorticity, which otherwise would not grow. We will compare the growth of the vorticity power spectrum from the vector modes of the measured stress tensor and see that it agrees with direct measurements of the vorticity power spectrum using the Delaunay method. Before we show this, we need to explain how we incorporate the measured stress tensor into the standard calculation of large scale evolution of density and velocity fields using perturbation theory.

\section{Large-Scale Corrections to PPF}
\label{PPFcorrect}

\subsection{Scalar-Vector Decomposition}

We are now ready to see the effects of orbit crossing in the large-scale evolution of density and velocity fields. We will use the equations of motion Eqs.~(\ref{continuity}-\ref{momentum}), supplemented by the Poisson equation,
\beq
\nabla^2 \phi = {3\over 2} {\cal H}^2 \Omega_m\, \delta,
\label{Poisson}
\eeq
and solve for the coupled system of $\delta$ and $\uv$, and treat the stress tensor as a {\em forcing} term, with known scale dependence and time evolution obtained from measurements in the simulations. 

A non-zero stress tensor leads to two new effects in the evolution of large-scale structure. We can decompose the velocity vector into scalar and vector modes, where the scalar mode is the velocity divergence (corresponding to the velocity parallel to the wave vector $\kv$) and the vector modes correspond to the vorticity (or the two components of $\uv$ perpendicular to $\kv$). In the pressureless perfect fluid (PPF) approximation, the divergence grows (since it couples to the gravitational potential), while any primordial vorticity decays in linear theory due to the expansion of the universe, since it does not couple to the gravitational force because it is conservative. Nonlinear terms can amplify vorticity but cannot create vorticity if there is no primordial contribution (as we assume in this paper). 

However, orbit crossing induces a nontrivial stress tensor and this will modify the evolution of the scalar and vector modes. From Eq.~(\ref{momentum}) we see that the new source for scalar and vector modes is the quantity
\beq
q_i(\xv,\tau) \equiv {1 \over \rho}\,  \nabla_j(\rho \sigma_{ij}).
\label{veldispforc}
\eeq 
More precisely we can decompose this into scalar and vector sources,
\beq
q_{\theta} \equiv \nabla\cdot\qv,\ \ \ \ \ 
\qv_w \equiv \nabla \times \qv,
\label{qsources}
\eeq
respectively. We will loosely call the correction from $\sigma_{ij}$ to the scalar modes (due to $q_\theta$)  ``velocity dispersion". Of course the velocity dispersion tensor $\sigma_{ij}$ affects the vector modes as well. However, in the simplest case of a diagonal $\sigma_{ij}$ which depends only on density, only $q_\theta$ survives. Note also that sometimes we refer to the stress tensor $T_{ij}=\rho \sigma_{ij}$ instead of the velocity dispersion tensor $\sigma_{ij}$ for conciseness. Finally, for simplicity we will refer to the decomposition of $\qv$ into $q_\theta$ and $\qv_w$ as ``decomposing the stress tensor into scalar and vector modes". 

Let us first write the linearized version of Eqs.~(\ref{continuity}-\ref{momentum}) taking into account Eqs.~(\ref{veldispforc}-\ref{qsources}). As usual, it is simplest to work with a different time variable,
\beq
\eta \equiv \ln D_+(\tau),
\label{eta}
\eeq
where $D_+$ is the linear growth factor, and scale the velocity and stress tensor so that,
\beq
\uv \rightarrow - {\cal H}f \, \uv, \ \ \ \ \ 
\sigma_{ij} \rightarrow ({\cal H}f)^2\, \sigma_{ij}, 
\label{scale}
\eeq
and thus also $q_\theta \rightarrow ({\cal H}f)^2 \, q_\theta$ and same for $\qv_w$.  Here $f=d\ln D_+/d\ln a\approx \Omega^{5/9}$ for $\Lambda$CDM models. Assuming that $f^2\approx \Omega_m$, the linearized equations of motion can be written after these transformations in the simple form,
\begin{eqnarray}
\partial_\eta\delta-\theta&=&0,
\label{eqn:continuity_2}\\   
\partial_\eta \theta+\frac{\theta}{2}-\frac{3\delta}{2} &=&
 \mbox{q}_\theta,
\label{eqn:momconservation_2}\\
\partial_\eta \wv+\frac{\wv}{2}&=& \qv_w.
\label{eqn:vortconservation_2}
\end{eqnarray}
\begin{figure}
\begin{center}
\includegraphics[width=75mm]{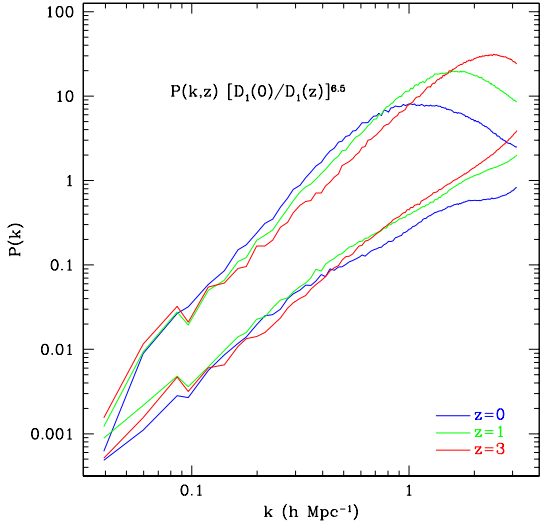}
\end{center}
\caption{Time dependence of the power spectra of the scalar and vector forcing terms $\mbox{q}_\theta$ (top three lines) and $\qv_w$ (bottom), see Eq.~(\protect\ref{qsources}). As in Fig.~\ref{a_dependence}, each power spectrum is linearly extrapolated to $z=0$. In this case, the time evolution of both forcing terms is fitted by $P(k,z)=[D_+(z)/D_+(0)]^{n_{vd}}P(k,0)$, with $n_{vd}=6.5\pm0.5$.} 
\label{a_dependence_veldisp}
\end{figure}
\begin{figure}
\begin{center}
\includegraphics[width=75mm]{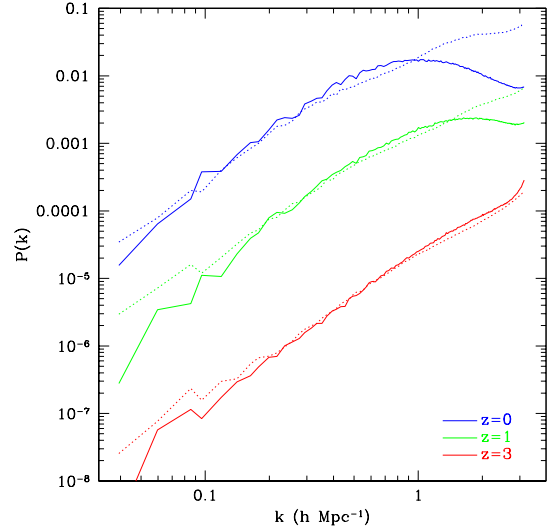}
\end{center}
\caption{Comparison between the Delaunay-estimated vorticity power spectrum (solid line) to the linear theory prediction from solving  Eq.~(\ref{vort_solution_PT}) (dotted line) for redshifts $z=0$, $z=1$  and $z=3$.} 
\label{vort_disp}
\end{figure}
Figure~\ref{a_dependence_veldisp} shows the power spectra corresponding to the two forcing terms and their time dependence, measured from the HR160 simulation using the method described in section~\ref{secveldisp}. Similarly to section~\ref{sec:a_dependence}, we fit for a time evolution of the form 
\beq
P_q(k,z) \propto [D_+(z)]^{n_{vd}},
\label{vortpower_a_dependence}
\eeq
where $P_q$ stands for both $P_{q_\theta}$ and $P_{q_w}$. We found that the best fit is $n_{vd}=6.5\pm0.5$, although the quality of the fit is not as good as in the case of the vorticity (see Fig.~\ref{a_dependence}). The reason for this may be shot-noise error coming from poorly sampled regions, where the error gets amplified by the $1/(1+\delta)$ factor in Eq.~(\ref{veldispforc}).

It is interesting to note that Eq.~(\ref{eqn:vortconservation_2}) provides us with a non-trivial consistency check between the vorticity power spectrum measured by the Delaunay method, and the adaptive method described in section~\ref{secveldisp} from which we measured the stress tensor and estimated the forcing term for vector modes. The vorticity and vorticity-forcing terms measured from the simulation should 
be consistent with the time evolution given by Eq.~(\ref{eqn:vortconservation_2}). Since this equation is decoupled from the other two equations (scalar and vector modes do not couple in linear theory), it can be easily solved. Ignoring the decaying mode, the {\em linear theory} solution for the vorticity power spectrum reads 
\beq
P_w(k)=\left(\frac{2}{n_{vd}+1}\right)^2 P_{q_w}(k).
\label{vort_solution_PT}
\eeq
Figure~\ref{vort_disp} shows the results of this consistency check. In it, we show the measured left and right hand sides of Eq.~(\ref{vort_solution_PT}) for redshifts $z=0,1,3$. The agreement in all cases is very good, improving, as expected, for higher redshifts.

\subsection{PT + Velocity Dispersion}
\label{extenddisp}

\begin{figure}
\begin{center}
\includegraphics[width=75mm]{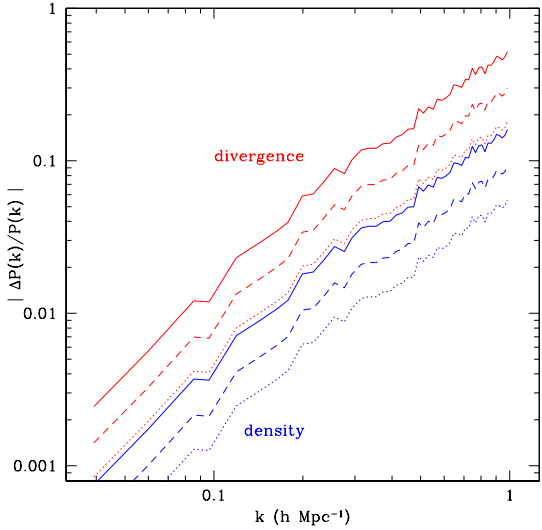}
\end{center}
\caption{Correction to the PPF approximation for the velocity divergence (three top lines) and density power spectrum (three bottom lines) due to velocity dispersion at redshifts $z=0$ (solid), $z=0.5$ (dashed) and $z=1$ (dotted). Note that the actual correction is negative in all cases, we plot their absolute values. These corrections are computed in linear theory, Eqs.~(\protect\ref{dispcorreq}) and (\protect\ref{tdispcorreq}), thus extrapolation well beyond $k \sim 0.1 \kMpc$ is only illustrative.} 
\label{dispcorrfig}
\end{figure}

We are interested in estimating the large-scale corrections to the PPF approximation due to the orbit-crossing induced $q_\theta$ and $\qv_w$. As we can see from the linearized equations of motion, Eqs.~(\ref{eqn:continuity_2}-\ref{eqn:vortconservation_2}), the scalar mode of the stress tensor corrects the PPF approximation already at the linear level, whereas the vector modes are decoupled in linear theory and correct the PPF at higher-order in PT. In this section we estimate the corrections due to the scalar mode $q_\theta$ (roughly speaking, velocity dispersion), while in the next section we tackle the corrections induced by $\qv_w$ at leading order in nonlinear PT. Since these deviations are small at large scales we can consider them separately.

The scalar mode correction can be included by writing the modified linear theory of Eqs.~(\ref{eqn:continuity_2}-\ref{eqn:momconservation_2}) in a compact form by using a two-component object $\psi_1=\delta$, $\psi_2=\theta$ that obeys the linear equations of motion,
\beq
\partial_\eta\psi_a(\kv,\eta)+\Omega_{ab}\ \psi_b(\kv,\eta)=Q_a(\kv,\eta),
\eeq
where $\Omega_{ab}$ is the 2x2 matrix,
\beq
\Omega_{ab}=\left(
    \begin{array}{cc}
      0 & -1 \\
      -\frac{3}{2} & \frac{1}{2} \\
    \end{array} \right) \\
\eeq
and $Q(\kv, \eta)=(0, \mbox{q}_\theta(\kv, \eta))$. The formal solution to these equations can be written as
\beq
\psi_a(\kv,\eta)=g_{ab}(\eta)\phi_b(\kv)+ \int_0^{\eta'} d\eta' g_{ab}(\eta-\eta')
Q_b(\kv,\eta'), 
\label{LinQ}
\eeq
where $\phi$ represents the initial conditions and $g_{ab}$ is the linear propagator~\cite{2006PhRvD..73f3520C},
\begin{equation}
g_{ab}(\eta)=\frac{e^\eta}{5}\left( 
  \begin{array}{cc}
     3 & 2\\
     3 & 2\\
  \end{array} \right) - \frac{e^{-3\eta/2}}{5}\left(
  \begin{array}{cc}
     -2 & 2\\
     3 & -3\\
  \end{array} \right)
  \label{gab}
\end{equation}
 Then, the density field in linear theory is given by
\beq
\delta(\kv,\eta)=\delta_{\rm ppf}(\kv,\eta)+
\frac{\mbox{q}_\theta(\kv,\eta)}{(n_{vd}/2-1)(n_{vd}/2+3/2)}, 
\eeq
where, as in Eq.~(\ref{vortpower_a_dependence}), we assumed that $\mbox{q}_\theta\propto D_+^{n_{vd}/2}$, and $\delta_{\rm ppf}(\kv,\eta)\equiv g_{ab}(\eta)\, \phi_b(\kv)$ is the usual linear theory evolved density field in the PPF approximation. We can then write the  density power spectrum to leading order in PPF corrections as
\beq
P_{\delta\delta}(\kv)=P_{\rm ppf}(\kv)+\frac{2\, P_{\delta \, q_\theta}(\kv)}{(n_{vd}/2-1)(n_{vd}/2+3/2)}
,
\label{dispcorreq}
\eeq
where $P_{\rm ppf}(\kv)$ is the linear density power spectrum in the PPF approximation, and $P_{\delta\, q_\theta}(\kv)$ is the cross power spectrum given by 
\beq
\langle \delta(\kv)\, q_\theta(\qv)\rangle=\delta_{\rm D}(\kv+\qv)\ P_{\delta\,  q_\theta}(\kv) ,
\label{Pdq}
\eeq
which we measure from the numerical simulations. From Eq.~(\ref{LinQ}) we can also obtain the velocity divergence in linear theory,
\beq
\theta(\kv,\eta)=\theta_{\rm ppf}(\kv,\eta)+
\frac{\mbox{q}_\theta(\kv,\eta)\, (n_{vd}/2)}{(n_{vd}/2-1)(n_{vd}/2+3/2)}, 
\eeq
and the corresponding power spectrum,
\beq
P_{\theta\theta}(\kv)=P_{\rm ppf}(\kv)+\frac{n_{vd}\, P_{\theta \, q_\theta}(\kv)}{(n_{vd}/2-1)(n_{vd}/2+3/2)}
.
\label{tdispcorreq}
\eeq
Note that the correction in the case of the velocity divergence power spectrum is a factor of $(n_{vd}/2\approx 3$) larger than in the case of the density; one can understand this from the fact that the divergence responds to the rate of change of the density fluctuations, and the correction to $\delta_{\rm ppf}$ grows as $D_+^{n_{vd}/2}$. It is also worth noticing that the corrections are negative, i.e. velocity dispersion tends to reduce the growth of structure; this is also expected for a stress tensor $\rho \sigma_{ij} \approx -p\, \delta_{ij}$ with a positive pressure (due to thermal motions) that is positively correlated with density fluctuations.  Figure~\ref{dispcorrfig} shows the absolute value of these corrections relative to the PPF approximation for both density and velocity divergence at redshifts $z=0,0.5,1$. We see that at $z=0$ the correction to the divergence power spectrum reaches 1\% at about $k\sim 0.1 \kMpc$, while for the density power spectrum this happens at about $k\sim 0.2 \kMpc$. By $z=1$ these scales shift by about a factor of two. At higher redshifts they rapidly decline as the growth factor changes rapidly before the onset of cosmic acceleration.

\begin{figure}[t!]
\begin{center}
\includegraphics[width=75mm]{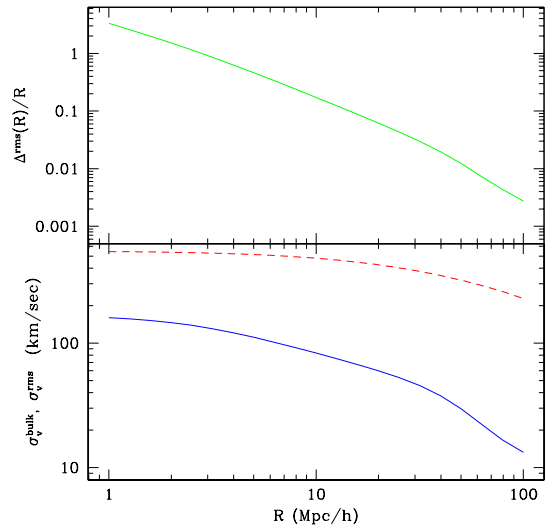}
\end{center}
\caption{{\em Top panel:} root mean square position fluctuations, Eq.~(\ref{posrms}), induced by velocity dispersion smoothed at scale $R$ divided by $R$. {\em Bottom panel:}  {\em rms} velocity dispersion, Eq.~(\ref{velrms}), in solid lines compared to {\em rms} bulk motions (dashed lines) smoothed on scale $R$. Note that velocity dispersion is smoothed on scales of order $1 \Mpc$, thus the solid line is an underestimate at small scales. All the quantities in this figure are evaluated at $z=0$.} 
\label{sigmatrace}
\end{figure}

These effects can be understood qualitatively, and to some extent quantitatively, by considering the typical size of the corrections to the velocities predicted by the single-stream, PPF approximation, smoothed over scale $R$. These corrections are,  in average, of order
\beq
\sigma_v^{rms}(R)\equiv{\cal H}f\ \left(\int d^3k\, P_\sigma(k)\, W^2_{\rm TH}(kR)\right)^{1/4},
\label{velrms}
\eeq
where $W_{\rm TH}(kR)$ is the Fourier transform of a top-hat filter of radius $R$, $P_\sigma(k)$ is the power spectrum of the trace of the velocity dispersion tensor (which is the dominant component), and the factor ${\cal H}f$ restores the correct units to $\sigma_{ij}$ (see Eq.~\ref{scale}). Equivalently, these velocity corrections can be interpreted as comoving position fluctuations, given by
\beq
\Delta^{rms}(R)\equiv\left(\int d^3k\, P_\sigma(k)\, W^2_{\rm TH}(kR) \right)^{1/4}.
\label{posrms}
\eeq
These two quantities are shown in Fig.~\ref{sigmatrace}. In the top panel, the ratio of the displacement corrections from Eq.~(\ref{posrms}) to the scale $R$ is plotted as a function of scale. An order of magnitude estimate of the effect on the density power spectrum can be obtained from the following argument. The dispersion in comoving positions given by $\Delta^{rms}(R)$ smooths out density perturbations. That suppression is approximately given by
\beq
P_{\rm smooth}(k)\sim P(k)\, {\rm e}^{-2\, (k \Delta^{rms}(2\pi/k))^2}.
\label{Ps}
\eeq
At large scales, e.g. $k\sim 0.1 \kMpc$, this gives a suppression consistent with the previously calculated density power spectrum corrections seen in Fig.~\ref{dispcorrfig}.

The bottom panel shows $\sigma_v^{rms}$ as a function of $R$, Eq.~(\ref{velrms}). We can see that the velocity dispersion on scales of $\sim100 \Mpc$ is of order 15 km/s. Comparing this dispersion with the single-stream bulk velocities on the same scale (dashed line), we conclude that the velocity dispersion corrections on those scales are small but, nevertheless, larger in relative terms than for the density power spectrum, in agreement with the detailed calculation presented in Fig.~\ref{dispcorrfig}.

In~\cite{2007PhRvD..75b1302A} it was argued that percent level corrections from orbit crossing to the density power spectrum are expected at $k \simeq 0.1 \kMpc$ based on a model of ``sticky dark matter". The effect discussed in that work is not an estimate of deviations from the PPF approximation, see Appendix~A in~\cite{2008PhRvD..77b3533C} for more details. Here we note that the estimate in~\cite{2007PhRvD..75b1302A} for the density power spectrum is two times larger than found here, and opposite in sign.

\subsection{PT + Vorticity}
\label{extendvort}

As discussed above, the effects of the vector modes of the stress tensor ($\qv_w$) on the density and divergence power spectrum only appear beyond linear theory, since in linear theory scalar and vector modes are decoupled. Here we estimate these corrections by calculating the one-loop density power spectrum including the effects from the vorticity, which is sourced by $\qv_w$. The strategy is as follows. We rewrite the equations of motion for density perturbations including now the nonlinear terms as follows,
\begin{widetext}
\beq
\partial_\eta{\delta}(\kv)-\theta(\kv)=\int d^3k_1 d^3k_2\,
\delta_D(\kv-\kv_{12})
\left[\frac{\kv\cdot\kv_2}{k_2^2}\delta_1\theta_2-
\frac{\kv_1\times\kv_2}{k_2^2}\cdot\delta_1
\wv_2\right], 
\label{eqn:continuity}
\eeq
\beq
\partial_\eta{\theta}(\kv)+\frac{\theta(\kv)}{2}-\frac{3\delta(\kv)}{2} =
\int d^3k_1 d^3k_2\delta_D(\kv-\kv_{12})
\left[
\frac{k^2(\kv_1\cdot\kv_2)\theta_1\theta_2}{2 k^2_1
  k^2_2}-\frac{(\kv_1\cdot\kv_2)(\kv_1\times\kv_2)\cdot
\theta_1\wv_2}{k^2_1 k^2_2}
\right],
\label{eqn:momconservation}
\eeq
\end{widetext}
where $\kv_{12}\equiv \kv_1+\kv_2$. To avoid cumbersome expressions, we have not written the time dependence of the fields explicitly. Also, on the right hand side, the subscripts ``1'' and ``2'' mean the fields evaluated at $\kv_1$ and $\kv_2$ respectively. 

A couple of points are worth noting. We have decomposed the velocity field into a divergence and a vorticity, and the latter will be taken as a known forcing term in the equations, since $\wv$ can be solved in linear theory as an uncoupled field from the measured $\qv_w$ (Eq.~\ref{eqn:vortconservation_2}) or directly measured from the Delaunay method.  In addition, note that we have neglected the $q_\theta$ source in the equation of motion for $\theta$, since this  effect was included already in the last section; here we are only interested in corrections due to $\qv_w$ alone, which enter through the $\wv$ forcing terms.

Following the compact notation introduced in the previous section, we can rewrite Eqs.~(\ref{eqn:continuity}-\ref{eqn:momconservation}) as 
\begin{widetext}
\beq
\partial_\eta \psi_a(\kv,\eta)+\Omega_{ab}\psi_b(\kv,\eta)=\int d^3k_1 d^3k_2\delta_D(\kv-\kv_{12})
[\gamma_{abc}(\kv_1,\kv_2)\psi_b(\kv_1,\eta)\psi_c(\kv_2,\eta)+ A_{ab}(\kv_1,\kv_2,\eta)\psi_b(\kv_1,\eta)],
\label{eqn:psi} 
\eeq
where $\gamma$ characterizes the non-linear mode coupling amplitudes and $A$ is the $\wv$-dependent forcing term. They can be written as 
\beq
\gamma_{112}=\frac{\kv_{12}\cdot\kv_2}{k_2^2}, \ \ \ \ \ 
\gamma_{222}=\frac{k_{12}^2(\kv_1\cdot\kv_2)}{2k_1^2 k_2^2}, \ \ \ \ \ 
A_{ab}= -\frac{\kv_1\times\kv_2}{k_2^2} \cdot\wv(\kv_2,\eta)
    \left( \begin{array}{cc}
      1 & 0\\
     0 &  \frac{\kv_1\cdot\kv_2}{k_1^2}
    \end{array} \right)
\eeq

Equation (\ref{eqn:psi}) can be formally solved by introducing again the linear propagator $g_{ab}$,  yielding:
\bea
\psi_a(\kv,\eta)&=&g_{ab}(\eta)\phi_b(\kv)+\int^\eta_0 ds\ g_{ab}(\eta-s)\int d^3k_1 d^3k_2\, \delta_D(\kv-\kv_{12}) \,
[\gamma_{bcd}(\kv_1,\kv_2)\psi_c(\kv_1,s)\psi_d(\kv_2,s)\nonumber \\ & & +A_{bc}(\kv_1,\kv_2,s)\, \psi_c(\kv_1,s)].
\label{eqn:psisolution}
\eea

This is an implicit form of the solution - it is written in terms of itself. However, it is suitable for a perturbative method. We write the field $\psi$ as a perturbative series: 
\begin{equation}
\psi_a(\kv,\eta)=g_{ab}(\eta)\phi_b(\kv) + \sum_{n=2} \psi_a^{(n)}(\kv,\eta),
\label{eqn:pertseries}
\end{equation}
and combining Eqs.~(\ref{eqn:psisolution}) and (\ref{eqn:pertseries}), we get a solution for the n-th order fields in terms of the lower order fields: 
\bea
\psi_a^{(n)}(\kv,\eta)&=&\int^\eta_0 ds\ g_{ab}(\eta-s)\int d^3k_1 \ d^3k_2 \  \delta_D(\kv-\kv_{12})
 \Big[\gamma_{bcd}(\kv_1,\kv_2) \sum_{r+s=n}  \psi_c^{(r)}(s,\kv_1)  \psi_d^{(s)}(s,\kv_2)
 \nonumber \\ & & +
A_{bc}(\kv_1,\kv_2,s)\, \psi_c^{(n-1)}(\kv_1,s)\Big] 
\eea
Thus, by knowing the linear solutions, we can calculate the solutions to any order. The linear solutions for $\delta$ and $\theta$ are just the PPF  linear solutions. In order to solve for the higher order fields $\delta$ and $\theta$, we need the vorticity field and its time dependence, which we have measured from dark matter N-body simulations.

As discussed above, the leading order correction due to vorticity effects appears in the one-loop contribution to the power spectrum. The density power spectrum, to that order, can be written as
\beq
P_\delta(\kv)\, \delta_D(\kv+\qv) = \langle\delta^{(1)}(\kv)\delta^{(1)}(\qv)\rangle+
 \left[\langle\delta^{(2)}(\kv)\delta^{(2)}(\qv)\rangle+ 2\langle\delta^{(1)}(\kv)\delta^{(3)}(\qv)\rangle\right]
\eeq
The first term is the usual tree-level (linear theory) power spectrum, while the terms in square brackets correspond to the one-loop correction, which are usually written as $P_{22}+P_{13}$ due to their dependence on PT order. Let us start by focusing on  the first one-loop term, which
leads to the $P_{22}$ contribution. The second order density can be written as  
\bea
\delta^{(2)}(\kv)&=&\int  d^3q_1 d^3q_2
\delta_D(\kv-\qv_{12}) \left[F(\qv_1,\qv_2)\delta^{(1)}
(\qv_1)\delta^{(1)} (\qv_2)+ \bm{G}(\qv_1,\qv_2)\cdot\wv(\qv_1)\delta^{(1)}(\qv_2)\right]\nonumber\\
&=&  \int  d^3q_1 d^3q_2 \delta_D(\kv-\qv_{12})\, [F_{12}\delta_1\delta_2+\bm{G}_{12}\cdot\wv_1\delta_2].
\eea
Note that in the second line we have just simplified the notation. The $P_{22}$ contribution to the power spectrum has four terms:
\bea
P_{22}(k)\, \delta_{\rm D}(\kv+\qv) &=&
 \int d^3q_1\ldots d^3q_4 \delta_D(\kv-\qv_{12}) 
\delta_D(\qv-\qv_{34})
\Big[F_{12}F_{34}\langle\delta_1\delta_2\delta_3\delta_4\rangle \nonumber \\ & &+ 
F_{12}\bm{G}_{34}\cdot\langle\delta_1\delta_2\wv_3\delta_4\rangle+
 \bm{G}_{12}F_{34}\cdot\langle\wv_1\delta_2\delta_3\delta_4\rangle+
G^\alpha_{12}G^\beta_{34}\langle w^\alpha_1\delta_2w^\beta_3\delta_4\rangle\Big] 
\eea
The first term is the usual PPF $P_{22}$ one-loop contribution. The second and third terms contain factors of the form $\langle\delta\wv\rangle$, which vanish due to symmetry. Then at large scales where connected contributions can be neglected we have
\beq
\langle w^\alpha_1\delta_2w^\beta_3\delta_4\rangle 
=\langle w^\alpha_1w^\beta_3\rangle\langle \delta_2\delta_4\rangle
=\delta_D(\qv_1+\qv_3)\,\delta_D(\qv_2+\qv_4) \ P^{\alpha\beta}_{ww}(\qv_1)P_\delta(\qv_2)
\eeq
from which we get the vorticity contribution to the $P_{22}$ power spectrum:
\beq
\Delta P_{22}(\kv)=\int d^3q\ 
G^\beta(-\qv,\qv-\kv)G^\alpha(\qv,\kv-\qv)\ P^{\alpha\beta}_{ww}(\qv)P_\delta(\kv-\qv)
=\int d^3q \frac{|\bm{G}(-\qv,\qv-\kv)|^2}{2}P_w(q)P_\delta(|\kv-\qv|),
\eeq
where we have used the actual vector structure of $\bm{G}(\kv,\qv)$ and the fact that
\beq
 P^{\alpha\beta}_{ww}(\qv)=\frac{P_w(q)}{2}\left[
   \delta_{\alpha\beta}-\frac{q_\alpha q_\beta}{q^2}\right].
\label{eqvortvectstruct}
\eeq

Similarly, we can compute the $P_{13}$ contribution to the density power spectrum. The third order density field can be written schematically as
\beq
\delta^{(3)}(\kv)=\int  d^3q_1 d^3q_2 d^3q_3
   \delta_D(\kv-\qv_{123})
   \big[
     H(\qv_1,\qv_2,\qv_3)\delta_1\delta_2\delta_3+
     \bm{R}(\qv_1,\qv_2,\qv_3)\cdot\wv_1\delta_2\delta_3+
     S^{\alpha\beta}(\qv_1\qv_2\qv_3)w^\alpha_1 w^\beta_2 \delta_3
\big].
\eeq
Then, the vorticity contribution to the power spectrum reads
\beq
\Delta P_{13}(k)=P_\delta(k) \int d^3q_1\,S^{\alpha\beta}(\qv_1,-\qv_1,-\qv)
P^{\alpha\beta}_{ww}(\qv_1)=P_\delta(k)\int d^3q\,S^{\alpha\alpha}(\qv,-\qv,-\kv)P_w(q),
\eeq
where in the last equality we have used Eq.~(\ref{eqvortvectstruct}). If one assumes that the time dependence of the vorticity is given by \mbox{$\wv\propto D_+^{n_w/2}$}, as found in section~\ref{secdelresults}, it is possible to write explicit expressions for the $\Delta P_{22}$ and $\Delta P_{13}$ power spectra: 
\beq
\Delta P_{22}(k)=\int d^3q P_w(q)
P_\delta(|\kv-\qv|)\frac{2k^2(1-x^2)}{q^2} \times 
\frac
{\left[(3+n_w)k^2+(1+n_w)q^2-4(1+n_w/2)\kv\cdot\qv \right]^2}
{n_w^2(5+n_w)^2|\kv-\qv|^2},
\label{expl_p22}
\eeq

\bea
\Delta P_{13}(k)&=& -P_\delta(k)\int d^3q  P_w(q) 
\frac{2(1-x^2)}{q^2|\kv+\qv|^2(5+n_w)(5+2n_w)}
\Big\{k^2\big[(3+n_w)(3+2n_w)k^2\nonumber \\ & & +(3+9n_w+2n_w^2)q^2\big]
+ 2(1+n_w)\, \kv\cdot \qv\, [q^2+(9+2n_w)k^2]+
4(2+3n_w/2)(\kv\cdot\qv)^2\Big\},
\label{expl_p13}
\eea
where $x$ is the cosine of the angle between $\kv$ and $\qv$. 

\end{widetext}

\begin{figure}
\begin{center}
\includegraphics[width=75mm]{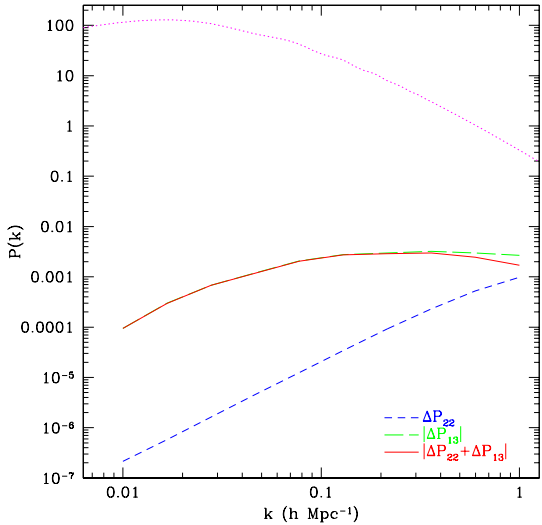}
\end{center}
\caption{Corrections to the density power spectrum  at $z=0$ due to stress tensor vector modes (vorticity effects), see Eqs.~(\ref{expl_p22}) and (\ref{expl_p13}). Note that the $\Delta P_{13}$ contribution (long dashed lines) is negative and larger in magnitude than the $\Delta P_{22}$ contribution (dashed lines). The total correction (solid lines) is negative and reaches 1\% of the {\em linear} spectrum (top dotted lines) at  $k\sim 1 \kMpc$, where further nonlinear effects not included here should become important.} 
\label{vort_corr}
\end{figure}

The end result of these calculations is that the leading large-scale contribution of vector modes of the stress tensor to the density power spectrum is fully specified in terms of the autocorrelation or power spectrum of the vorticity, which we have measured from the simulations. Figure~\ref{vort_corr} shows the results of these calculations. We see that the total correction is negative, as expected physically, and very small at large scales. For example, at $k\sim 0.1 \kMpc$ where the scalar modes contributed percent level corrections,  the modifications of the PPF approximation from vector modes is about $10^{-4}$ of the linear spectrum, and thus totally negligible. The reason for this is that by symmetry the vorticity does not couple to the scalar modes, it is only through vorticity squared that the effect is present. We expect similar results for the velocity divergence power spectrum within a factor of a few, still completely negligible at large scales.

\section{Conclusions}
\label{conclude}

We studied the impact of orbit crossing in the large-scale power spectra of density and velocity divergence fields, which are usually described in the pressureless perfect fluid (PPF) approximation. We presented a method to  extend perturbation theory (PT) beyond the PPF approximation, based on measuring the stress tensor induced by orbit crossing in numerical simulations. The stress tensor, when decomposed into scalar and vector modes leads to corrections associated with velocity dispersion and the effects of vorticity. We found the effects due to the scalar modes to be small, but not negligible at large scales ($k\simeq 0.1 \kMpc$), particularly for the velocity divergence power spectrum (see Fig.~\ref{dispcorrfig}). The impact of vorticity on large scales is much smaller, see Fig.~\ref{vort_corr}. These two effects appear at different orders in PT and have been included separately as we are interested in large scales where the induced corrections are small. Both lead to suppressions of the power spectra predicted by the PPF approximation, as expected physically since velocity dispersion and vorticity should inhibit collapse. In this regard we emphasize that neglecting orbit crossing has {\em opposite} effects on Eulerian compared to Lagrangian PT. For Lagrangian PT, neglecting orbit crossing leads to (much more severe) {\em underestimates} of the density power spectrum (see e.g.~\cite{2007arXiv0711.2521M} for a recent example), since neglecting self-gravity in caustics leads to artificial thickening of such structures when trajectories cross without interacting. 

A novel aspect of our calculation is the estimation of the stress tensor and the vorticity and divergence power spectra from numerical simulations. To estimate velocity fields, we applied the Delaunay tessellation method, which we have shown to be a more reliable estimator than traditional mass weighting schemes. While estimates of the velocity divergence are robust, we found that measurements of the vorticity power spectrum are significantly more difficult, due to aliasing during the measurement process and most importantly lack of resolution in the simulations. For the latter we have found that low resolution simulations can overestimate the vorticity power spectrum by an order of magnitude. This maybe be due to insufficient spatial resolution in multistreaming regions, with the overestimate perhaps related to aliasing effects during the PM part of the force calculation, which may generate a vector mode. In any event, for high enough resolution we find that the vorticity power spectrum converges to a stable answer. On the other hand, care must be taken that these spurious effects are not present when using numerical simulations to study nonlinear velocities, since artificial vorticity can amplify the velocity power spectrum at small scales.

A nontrivial check of our numerical calculation of the stress tensor, which we have done using an adaptive method independent of the Delaunay tesselation, is that its vector modes source the growth of vorticity. Therefore, using linear PT from this vector source one should recover at large scales the vorticity power spectrum measured by the Delaunay method, as we do (see Fig.~\ref{vort_disp}). This does not test the scalar mode of the stress tensor though, which ends up inducing the largest correction to the PPF approximation. In this respect, it would be interesting to test how robust the scalar part of the stress tensor is to details of the numerical simulations, as spurious effects due to discreteness may amplify velocity dispersion in simulations~\cite{1997ApJ...479L..79M,1998ApJ...497...38S} (see also~\cite{2008arXiv0805.1357J}). As far as we know, our work is the first to make a quantitative connection between the growth of velocity dispersion and that of the density power spectrum, which will be useful to probe more in order to make sure that simulations can correctly reproduce the matter power spectrum to percent level, as required for the next generation of weak lensing surveys designed to probe cosmic acceleration~\cite{2005APh....23..369H}.

The deviations we found from the PPF approximation at large scales are small but not negligible, in particular for the velocity divergence power spectrum, for which corrections are a factor of about three larger than for the density power spectrum. Our estimate, being based on numerical simulations, corresponds to fixed cosmological parameters (e.g. $\sigma_8=0.9$, $\Omega_m=0.27$ and $n_s=1$). Given the strong dependence on the growth factor of the correction ($\propto D_+^{2.25}$ relative to PPF) we expect it to be smaller for lower normalization amplitudes, as well for cosmological parameters that correspond to less power at small scales (i.e. $\Omega_m<0.27$ and $n_s<1$). 

In section~\ref{orbitcross} we sketched what must be done to include these effects from first principles into analytic calculations such as RPT that usually start from the PPF approximation, instead of using the hybrid approach we develop here partially based on numerical simulations.  Including velocity dispersion in a self-consistent manner should cure divergences that appear in PT for scale-free models with initial power spectra $P(k)\sim k^n$ for $n>-1$, as well as regulate the divergences that appear in the resummation of the Lagrangian space propagator~\cite{2008arXiv0805.0805B}. In addition, we expect velocity dispersion and vorticity to be crucial to describe the virial turnover in the density power spectrum.  Another interesting application of the ideas presented in section~\ref{orbitcross} is to use the cumulant hierarchy to describe nonlinear effects in a massive neutrino component, to improve on recent calculations~\cite{2008PhRvL.100s1301S,2008arXiv0809.0693W} that assume linearity. We hope to report on some of this in the near future.

\acknowledgements

We thank M.~Blanton, M.~Crocce, A.~Gruzinov, A.~McFadyen P.~McDonald, E.~Schucking, L.~Senatore and R.~Sheth for useful discussions. This work was partially supported by NSF AST-0607747 and NASA NNG06GH21G. S.~P.~ was partially supported by a James Arthur Graduate Assistantship at NYU. R.S. thanks the Aspen Center for Physics and the Physics and Astronomy Department at the University of Pennsylvania for hospitality while this work was being finished.

\appendix

\section{Estimating the Velocity Field}
\label{secdelaunay}

\subsection{The Delaunay Tessellation}

One of the main obstacles in measuring the velocity field from cosmological simulations is the fact that it is only sampled on a
discrete set of points. One encounters the same difficulty when reconstructing the peculiar velocity field from observations, where
velocity is only sampled at the locations of galaxies.  One can identify two problems associated with this fact. On one hand, since the velocity is only known at points where the mass is located, almost all procedures to reconstruct the velocity field from a discrete sample give mass-weighted quantities, while most theoretical predictions concern volume-weighted quantities. On the other hand, low density regions are very sparsely sampled, and therefore subject to large Poisson errors. Laying out a structured grid, as it is often done, leaves grid points empty in such regions with the consequence that the velocity field is set to zero (thus missing outflows in voids), while in practice the velocity field is undetermined due to the poor mass resolution. These issues have been recognized for a long time in the theoretical large-scale structure literature, see e.g.~\cite{1995ApJ...442...39J,1996MNRAS.279..693B,1997MNRAS.290..566B}.

The work in~\cite{1996MNRAS.279..693B} introduced two new velocity estimation methods that attempt to overcome these problems. These methods are based on the Voronoi and Delaunay tessellations of the discrete set of points where the velocity field is sampled. They showed that the Delaunay Tessellation method has fewer computation requirements than the Voronoi Tessellation method, while giving equally or more reliable results. Thus, in our work we will consider only the Delaunay method, following~\cite{1996MNRAS.279..693B} to a large extent. Our implementation of the method is based on the public code {\sf Qhull}~\cite{Qhullref} to construct the Delaunay tessellation. Many other applications of Delaunay tessellations have recently appeared in the large-scale structure literature (see e.g.~\cite{2000A&A...363L..29S,2007MNRAS.382....2R}), see also~\cite{2007arXiv0708.1441V} for an in-depth review and other applications.

\begin{figure*}
\begin{center}
\includegraphics[width=120mm]{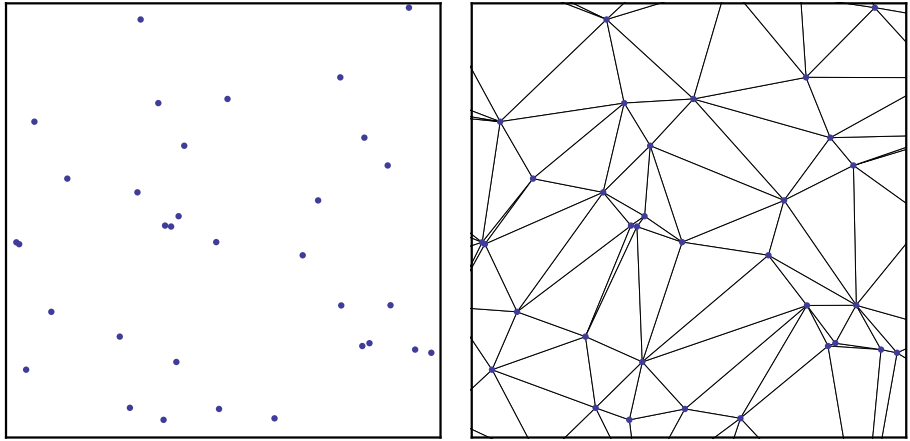}  
\end{center}
\caption{Delaunay tessellation of a set of random points in two dimensions. The left panel shows the original set of points. In the right panel, we show the regions corresponding to the tessellation. Note that in two dimensions the Delaunay tessellation consists of triangles instead  of tetrahedrons.}
\label{delaunay_intro}
\end{figure*}

The formal definition of the Delaunay tessellation ${\cal D(P)}$ of a set of points ${\cal P}$ (in three dimensions) is the set of tetrahedrons defined by four points whose circumscribing sphere is {\em empty} in the sense that no point of the generating set ${\cal P}$ should be inside the circumsphere~\cite{Delaunay1934}. In Fig.~\ref{delaunay_intro} we show an example of the tessellation of a set of random points in two dimensions.  It can be shown that the Delaunay tessellation is unique. Moreover, the Delaunay tetrahedrons are objects of minimal size and elongation. These characteristics make the Delaunay method optimal for a three dimensional interpolation.

\subsection{Reconstructing the Velocity Field}

Once the Delaunay tessellation from the set of sample points is obtained, it is possible to estimate the velocity at each point $p$ in space by linearly interpolating the velocities at the vertices of the tetrahedron that contains the point $p$. This procedure leads to a continuous velocity field with constant gradient in each tetrahedron.

Mathematically, we can express the Delaunay method to find the velocity $\uv$ at a point $p$ with coordinates $\xv$ as follows. Let $\xv_i$, with $i=0,1,2,3$, be the coordinates of the vertices of the tetrahedron containing $p$. Since the Delaunay tetrahedrons are non-degenerate (i.e. they do not collapse into 2D objects), we can express $\xv$ as a linear combination of $\xv_i$:
\beq
\Delta\xv=\sum_{i=1}^3\alpha_i\Delta\xv_i,
\label{eqdelx}
\eeq
where $\Delta\xv\equiv\xv-\xv_0$ and $\Delta\xv_i\equiv\xv_i-\xv_0$. The linear interpolation of the velocity at point $p$ is simply 
\beq
\Delta\uv=\sum_{i=1}^3\alpha_i\Delta\uv_i,
\label{eqdelu}
\eeq
where  $\Delta\uv\equiv\uv-\uv_0$, $\Delta\uv_i\equiv\uv_i-\uv_0$, and $\alpha_i$ satisfy Eq.~(\ref{eqdelx}). Thus, the problem reduces to solve for the $\alpha_i$, which can be readily be written as
\beq
\left(\begin{array}{c}
  \alpha_1\\
  \alpha_2\\
  \alpha_3
\end{array} \right) =
{\mathbf A}^{-1} \cdot 
\left(\begin{array}{c}
  \Delta x\\
  \Delta y\\
  \Delta z
\end{array} \right),
\eeq
where $\Delta x$, $\Delta y$ and $\Delta z$ are the components of $\xv$, and $\mathbf A$ consists of the components of the $\Delta\xv_i$:
\beq
{\mathbf A}\equiv
\left(\begin{array}{c c c}
  \Delta x_1 & \Delta x_2 & \Delta x_3 \\
  \Delta y_1 & \Delta y_2 & \Delta y_3 \\
  \Delta z_1 & \Delta z_2 & \Delta z_3 \\
\end{array}\right).
\label{eqdefA}
\eeq

These equations allow us to compute an estimation of the velocity field for any point in the simulation volume. In particular, we are interested in determining the volume-averaged field at a given set of points, often a grid. Typically, one wants to compute the average
$\uv_R(\rv_i)$ of the velocity field in spheres of radius $R$ centered at the points $\rv_i$, usually on a grid. In order to obtain that, one can carry out the following algorithm~\cite{1996MNRAS.279..693B}:
\begin{enumerate}
\item{Construct the Delaunay tessellation from the locations of the  simulation particles.}
\item{For each point $\rv_i$:}
\begin{enumerate}
  \item{Find the intersection of the Delaunay tetrahedrons with a sphere  of radius $R$ centered at $\rv_i$.}
  \item{For each intersecting tetrahedrons $j$, determine intersection  volume $V_j$ and mean velocity $\uv_j$ in that volume.}
  \item{Compute $\sum_j\,V_j\uv_j/(4\pi R^3/3)$. }
\end{enumerate}
\end{enumerate} 

However, both constructing the tessellation for the large number of
particles \mbox{($\sim 10^9$)} typical of state-of-the-art
simulations and finding the tetrahedrons intersecting a given sphere
are very time-consuming. For the sake of efficiency, we modified the
previous procedure. Instead of calculating the velocity average on a
sphere centered at a grid point $\rv_i$, we compute the average in the
volume given by all Delaunay tetrahedrons {\em totally contained} in
the grid cell corresponding to $\rv_i$. Thus, our approximate
algorithm reads as follows: 
\begin{enumerate}
\item{For each grid point $\rv_i$,}
\begin{enumerate}
  \item{Construct the Delaunay tessellation of the points {\em contained}
  in the corresponding grid cell.}
  \item{Compute the volume $V_j$ and mean velocity $\uv_j$ for every
  Delaunay tetrahedron.}
  \item{Compute $\sum_j\,V_j\uv_j/\sum_j\,V_j$.}
\end{enumerate}
\end{enumerate}
It is possible to write explicit expressions for the volume $V_j$
and mean velocities $\uv_j$ in a Delaunay tetrahedron. It follows from
elementary geometry that:
\beq
V_j=|\det({\mathbf A})|,\,\,\,\,\,\,\,\,\,
\uv_j=\frac{1}{4}\sum_{i=0}^3 \uv_j^{(i)},
\eeq
where ${\mathbf A}$ is defined in Eq.~(\ref{eqdefA}), and $\uv_j^{(i)}$ are the four velocities at the vertices of the tetrahedron $j$.

Note that in the new algorithm, we only construct the tessellation of
the points inside the grid cell, a much smaller number of particles
than the total simulation. Moreover, it is no longer necessary to find
the tetrahedrons or their intersection with a sphere. Nevertheless, with
this procedure, the average volume will in general vary from grid point
to grid point. We reduce this undesired effect by only using relatively
coarse grids, where we expect a more uniform distribution of
particles. We are thus obtaining a smoothed field
$\uv_R(\xv)$ given by
\beq
\uv_R(\xv)=\int d^3y\, W_R(\xv-\yv)\, \uv(\yv),
\label{delaunay_kernel}
\eeq
where $W_R(\xv)$ is a spherical top-hat filter with $R \approx (3V_{\rm cell}/4\pi)^{1/3}$.  To deconvolve Fourier space quantities, we thus divide by the Fourier transform of $W_R$. Note that this correction is only correct on average, tests reveal that it is accurate to about 1\% at $k=0.2 \kMpc$, more than enough for our purposes in this paper, but not enough for precision tests of velocity divergence power spectra in the weakly nonlinear regime.

\begin{figure*}[ht!]
\begin{center}
\includegraphics[width=0.49\textwidth]{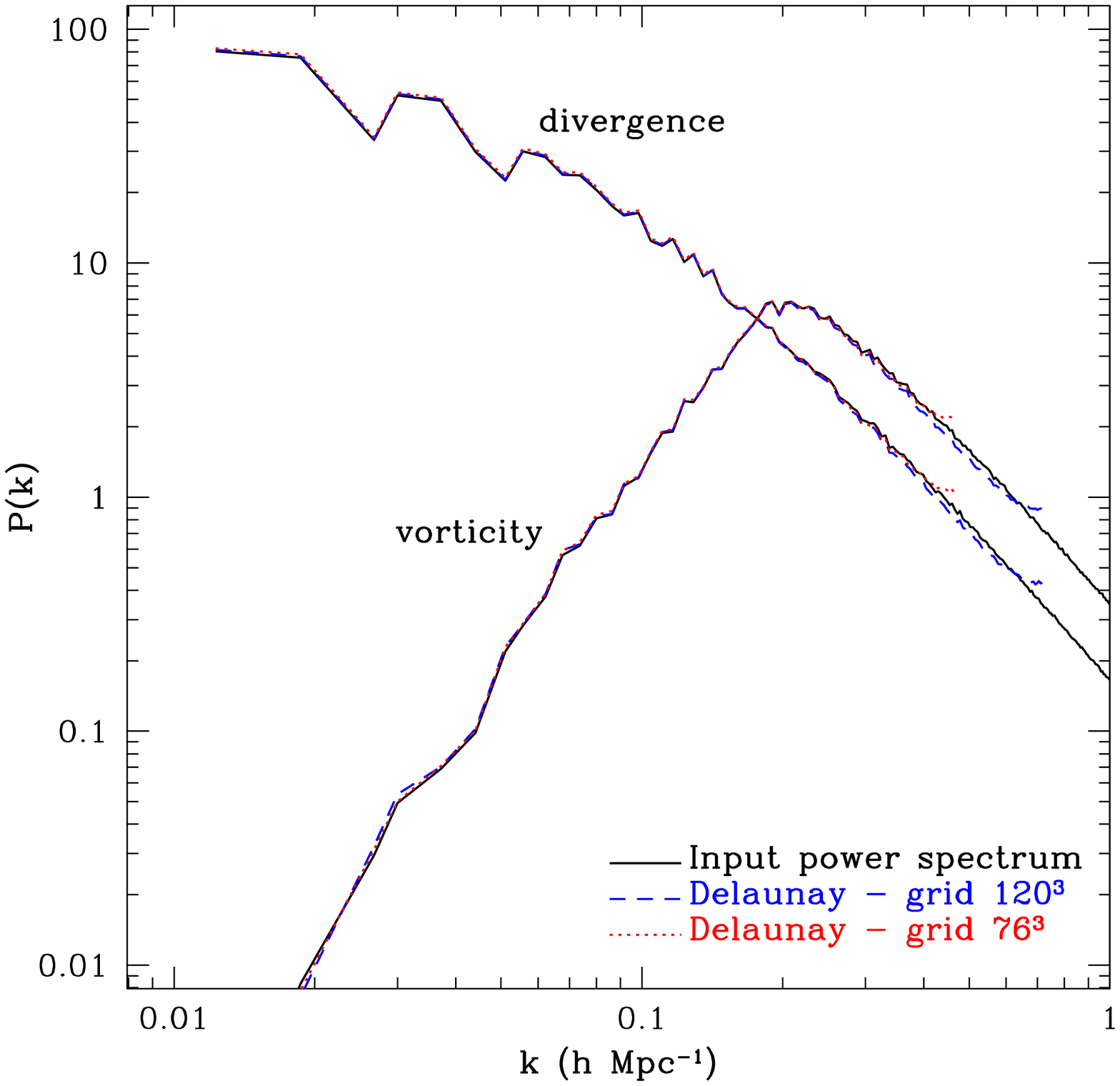}
\includegraphics[width=0.49\textwidth]{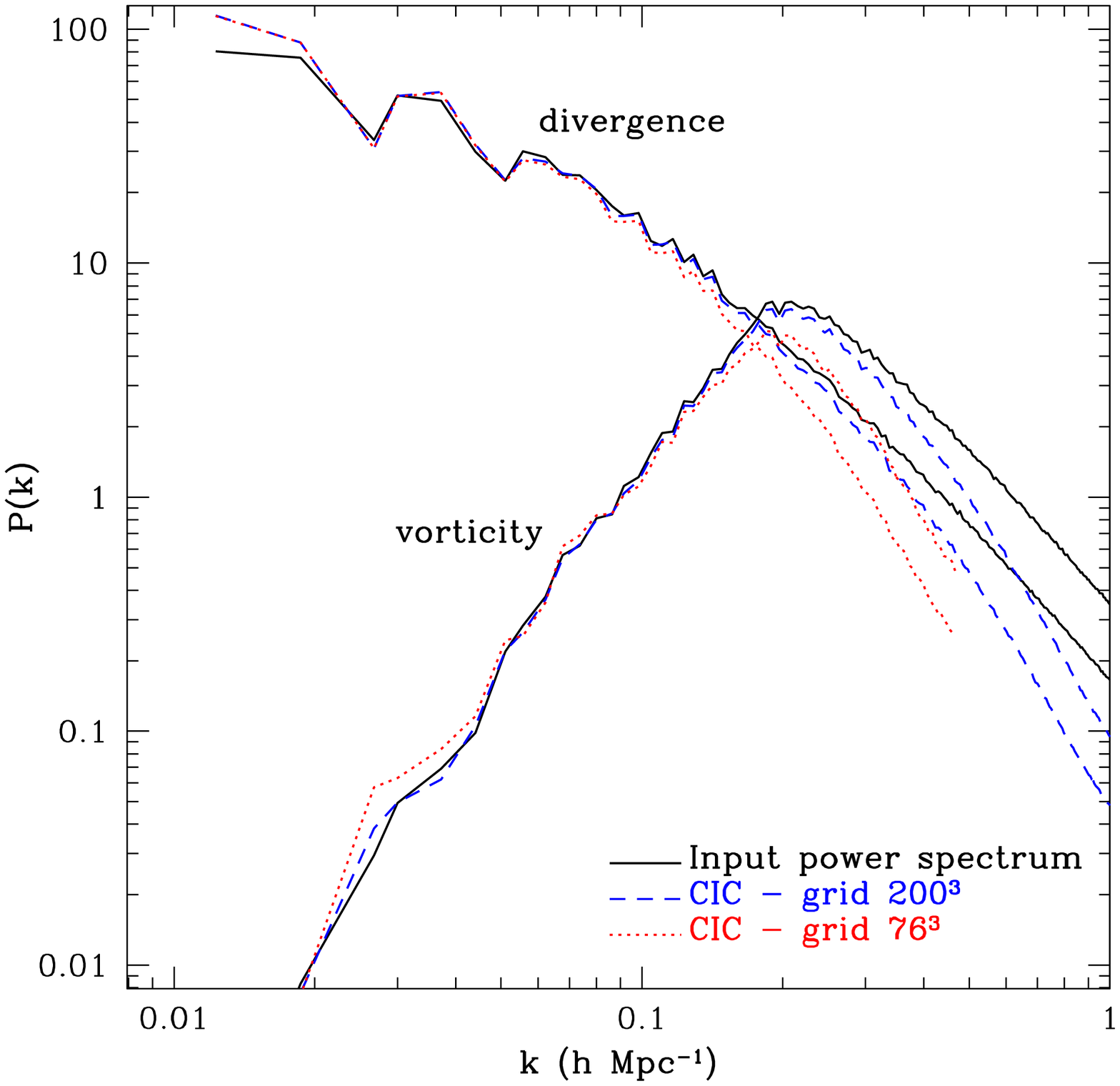} 
\caption{{\em Left panel:} Delaunay estimated divergence and vorticity power spectra. The solid lines correspond to the power spectra used to generate the velocity field. The dashed lines are the spectra measured on a $120^3$ grid, and the dotted lines are measured on a $76^3$ grid. All spectra are corrected by deconvolving the smoothing kernel (Eq.~\protect\ref{delaunay_kernel}). {\em Right panel:} Same as left panel, but using the CIC method. Note that since these measured spectra are obtained as the ratio of two interpolated quantities, they cannot be easily corrected for the window of the interpolation scheme.}
\label{delaunay_test}
\end{center}
\end{figure*}

\subsection{Testing the Delaunay Method}

One of the difficulties of testing accuracy of the Delaunay method is that it is expected to be more accurate (in measuring volume-weighted quantities) than the traditional estimations from mass-weighted schemes. Thus, we lack a more trustworthy method to use as reference. Out strategy to overcome this difficulty is setting up a ``controlled numerical experiment'': we generate a random Gaussian velocity field with given divergence and vorticity power spectra and then use the Delaunay method to recover the velocity statistics. In this way, we can compare the results of the method with the exact input power spectra used to generate the velocity field. 

For the sake of comparison, we also measure the velocity power spectra with the well known Cloud-in-Cell mass-weighted method (CIC). It consists of interpolating the particles mass and velocity on a grid using the CIC kernel $W_{\rm CIC}(\xv)\equiv\prod_i W_{\rm CIC}(x_i)$ defined by
\beq
W_{\rm CIC}(x_i)= \left\{\begin{array}{ll}
    1-|x_i|&\,\,\,\,\mbox{for}\,|x_i|<1\\
    0      &\,\,\,\,\mbox{for}\,|x_i|\geq 1,
  \end{array} \right.
\eeq
where $\xv$ is measured in units of grid separation. Note that interpolating the particle velocities by using this method gives the momentum field instead of the velocity field. Thus, one needs to compute the ratio between this quantity and the density field to obtain the velocity. As we mentioned above, in underdense regions if the grid is made too fine there will be grid points for which no particle is assigned, which means there is no information on the velocity field, but typically one would set to zero (incorrectly) the velocity. In addition, dividing the interpolated momentum by the interpolated density means that it is difficult to correct for the interpolation kernel after the velocity field is Fourier transformed, unlike the case of the density field (see~\cite{2008arXiv0804.0070C} for a recent discussion of interpolation corrections and comparison of CIC with other mass assignment schemes for the density field).

We generate a Gaussian velocity field on a grid of $400^3$ cells with a  divergence and vorticity power spectra based roughly on expectations from previous measurements in the literature~\cite{2001NYASA.927...13S}  and then interpolate this velocity on the positions of the $640^3$ dark  matter particles obtained from running an N-body simulation with  {\sf Gadget2}~\cite{2005MNRAS.364.1105S} (see Table~\ref{tablesimulationsdel} below for more details on the simulations).  Then we measure the divergence and vorticity power spectra using the Delaunay method and the CIC method.

The results are shown in Fig.~\ref{delaunay_test}.  We applied the Delaunay method, as described in the previous subsection, on a coarse grid of $76^3$ cells and a finer grid of $120^3$ cells, and measured the power spectra using fast Fourier transforms. On scales close to the Nyquist frequency, the power spectra were corrected by deconvolving the kernel defined in Eq.~(\ref{delaunay_kernel}). The recovered divergence and vorticity agree very well with the input power-spectra (left panel).  In the right panel,  we show the results of the CIC method. We see that in order to obtain results comparable to the Delaunay method, one needs to use a much finer interpolation grid. Even in that case, there are considerable differences on large scales. On small scales, the power spectrum is underestimated. This is due to the CIC interpolation kernel. However, it cannot be corrected as in the Delaunay case because the velocity field was obtained as the ratio of two CIC-interpolated quantities. In principle, one could deconvolve the density and the momentum interpolated fields before taking the ratio, but this procedure does not give good results because it introduces noise in the deconvolved fields~\cite{2004PhRvD..70h3007S}. In particular, we observed that the deconvolved density field has a non-negligible number of negative density grid points. Similar results on the better noise properties of the Delaunay method were obtained for the PDF of the velocity divergence in~\cite{1996MNRAS.279..693B}.

\subsection{Sampling Effects: Aliasing}
\label{secaliasing}

One of our goals in this paper is to estimate reliably the vorticity field from cosmological simulations. Since at large scales it is expected to be very small compared to the divergence, it is important to analyze the {\em sampling effects}. We will show that such effects make the sampled vorticity field a mixture of the vorticity and divergence of the original field.

Let us assume we know the velocity field $\uv(\xv)$ inside a box of volume $L^3$, and will study the effects of sampling that field on a grid of $N^3$ cells. Let us decompose the original (exact) field $\uv(\xv)$ in Fourier series:
\beq
\uv(\xv)=\sum_\kv \bar\uv(\kv)e^{i\kv\cdot\xv},
\eeq
where the vector $\kv$ has components $k_\alpha=2\pi n_\alpha/L$ with $n_\alpha\in \mathbb{Z}$, i.e. arbitrarily large  frequencies appear in the Fourier sum. We want to compare these exact Fourier modes $\bar\uv(\kv)$ to the discrete Fourier modes $\hat\uv(\qv)$ on the
grid ($q_\alpha=2\pi m_\alpha/L$,\ $m_\alpha\in\mathbb{Z}$,\ $0\leq m_\alpha<N$). We can write the discrete Fourier transform of the
velocity field as 
\bea
\hat\uv(\qv)&=&\frac{1}{N^3}\sum_{\xv}\uv(\xv)e^{-i\qv\cdot\xv}
\nonumber\\ 
&=&\frac{1}{N^3}\sum_\xv\sum_\kv\bar\uv(\kv)e^{i(\kv-\qv)\cdot\xv},
\eea
where $\qv$ and $\xv$ are on the grid. This can be further simplified 
into
\beq
\hat\uv(\qv)=\sum_\kv \frac{\bar\uv(\kv)}{N^3} \prod_{\alpha=1}^3 
\frac{1-\exp[iL(k_\alpha-q_\alpha)]}
     {1-\exp[\frac{iL}{N}(k_\alpha-q_\alpha)]}.
\eeq
The product vanishes unless $\rv\equiv\kv-\qv=\frac{2\pi N}{L}\mv$
with $\mv$ an integer vector. Finally, we obtain 
\beq
\hat\uv(\qv)=\sum_\rv \bar\uv(\qv+\rv),\,\,\,\,\,\,
\rv=\frac{2\pi N}{L}\mv.
\eeq
This equation states explicitly that velocity Fourier modes beyond
the Nyquist wavenumber of the grid ($k_{Ny}=\pi N/L$) affect the
grid Fourier modes. This is known as {\em aliasing}. To see how
this effect appears in the vorticity power spectrum, let us assume
that the velocity field is purely potential, that is,
$\bar\uv(\kv)=i \kv \theta(k)/k^2$. Evidently, 
\mbox{$\bar{\mathbf w}(\kv)\equiv i \kv\times\bar\uv(\kv)=0$}, but
\beq
\hat{\mathbf w}(\qv)\equiv i \qv\times\hat\uv(\qv)=
\sum_\rv \frac{i\qv\times\rv}{|\qv+\rv|^2}\theta(\qv+\rv)
\eeq
does not vanish. Moreover, the sampled vorticity power spectrum can be
written as
\beq
\hat{P}_w(\qv)=\sum_\rv\frac{|\qv\times\rv|^2}
{|\qv+\rv|^4}P_\theta(|\qv+\rv|), 
\label{eqaliasing}
\eeq
where $\rv=(2\pi N/L)\mv$ denotes multiples of the Nyquist frequency. Note that $P_\theta$ is the power spectrum of the {\em original}
divergence field. 

\begin{figure}
\begin{center}
\includegraphics[width=75mm]{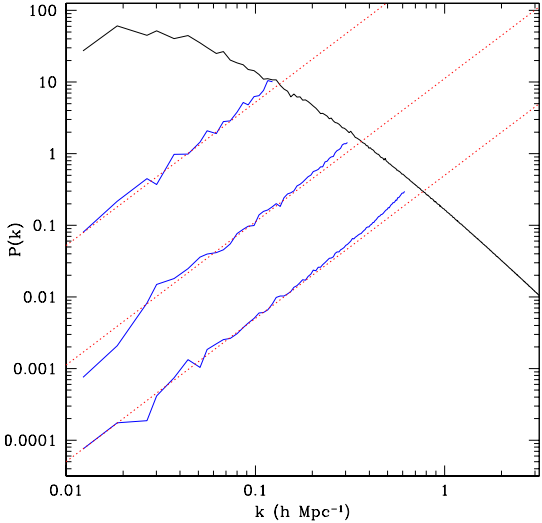}
\end{center}
\caption{Measured divergence and vorticity power spectra from a {\em vorticity-free} velocity field sampled on three different grids of $20^3$, $50^3$ and $100^3$ grid points (top to bottom). This shows  that the measured vorticity is purely due to aliasing. The dotted lines correspond to the predictions of Eq.~(\protect\ref{eqaliasing}).}
\label{novor}
\end{figure}

Equation~(\ref{eqaliasing}) tells us that the velocity divergence of wavenumbers larger than the Nyquist of the grid induces a spurious
vorticity in the sampled field. In the low-q limit ($q\ll 2\pi N/L$), it is easy to see that $\hat{P}_w(\qv)\propto q^2$. Figure~\ref{novor} shows the predictions of the large scale limit of this formula. We generated a zero-vorticity velocity field, which was sampled {\em without smoothing it} on different grids. Then, we compared the measured vorticity power spectrum to the predictions of Eq.~(\ref{eqaliasing}). Note that the amplitude of the correction was not fitted: the formula estimates correctly both  the low-$k$ limit dependence of the vorticity power spectrum and the amplitude of the spurious vorticity.

It is important to remark that we cannot directly extrapolate these results to the vorticity estimates of Fig.~\ref{delaunay_test}. In
that case, both the vorticity and divergence fields are smoothed on a scale given by the grid separation. This procedure greatly reduces the power spectrum on wavenumbers larger than the Nyquist frequency of the grid, making the aliasing effect less important. However, this analysis is useful to set an upper-bound estimate of aliasing effects.

\bibliography{masterbiblio}

\end{document}